\pdfoutput=1
\documentclass[fleqn,usenatbib]{mnras}

\usepackage[T1]{fontenc}
\usepackage{ae,aecompl}

\usepackage{soul,color,xcolor}
\usepackage{hyperref}
\usepackage{graphicx}
\usepackage{amsmath}
\usepackage{amssymb}
\usepackage{widetext}
\usepackage{bm}
\usepackage{lineno}
\usepackage{lastpage}
\usepackage[absolute,overlay]{textpos}
\usepackage{newtxtext,newtxmath}
\usepackage[section]{placeins}
\usepackage{float}

\newcommand{\md}{\mathrm{d}}
\newcommand{\ej}{_\mathrm{ej}}
\newcommand{\kgrey}{\kappa_\mathrm{grey}}
\definecolor{cyan}{rgb}{0.0, 1.0, 1.0}

\title[Grey-body kN Population]{Modelling Populations of Kilonovae}

\author[C. N. Setzer et al.]{Christian N. Setzer$^{1}$\thanks{E-mail: christian.setzer@fysik.su.se},
Hiranya V. Peiris$^{1,2}$,
Oleg Korobkin$^{3,4,5}$,
Stephan Rosswog$^{6}$
\\
$^{1}$The Oskar Klein Centre, Department of Physics, Stockholm University, AlbaNova, SE-10691 Stockholm, Sweden\\
$^{2}$Department of Physics and Astronomy, University College London, Gower Street, London WC1E 6BT, UK\\
$^{3}$CCS-7, Los Alamos National Laboratory, P.O. Box 1663, Los Alamos, NM 87545, USA\\
$^{4}$Center for Theoretical Astrophysics, Los Alamos National Laboratory, Los Alamos, NM 87545, USA\\
$^{5}$Joint Institute for Nuclear Astrophysics - Center for the Evolution of the Elements, USA\\
$^{6}$The Oskar Klein Centre, Department of Astronomy, Stockholm University, AlbaNova, SE-10691 Stockholm, Sweden
}

\date{Accepted XXX. Received YYY; in original form ZZZ}

\pubyear{2022}

\begin{document}
\label{firstpage}
\pagerange{\pageref{firstpage}--\pageref{LastPage}}
\maketitle


\begin{abstract}
  The 2017 detection of a kilonova coincident with gravitational-wave emission has identified neutron star mergers as the major source of the heaviest elements, and dramatically constrained alternative theories of gravity. Observing a population of such sources has the potential to transform cosmology, nuclear physics, and astrophysics. However, with only one confident multi-messenger detection currently available, modelling the diversity of signals expected from such a population requires improved theoretical understanding. In particular, models which are quick to evaluate, and are calibrated with more detailed multi-physics simulations, are needed to design observational strategies for kilonovae detection, and to obtain rapid-response interpretations of new observations. We use grey-opacity models to construct populations of kilonovae, spanning ejecta parameters predicted by numerical simulations. Our modelling focuses on wavelengths relevant for upcoming optical surveys, such as the Rubin Observatory Legacy Survey of Space and Time (LSST). In these simulations, we implement heating rates that are based on nuclear reaction network calculations. We create a Gaussian-process emulator for kilonova grey opacities, calibrated with detailed radiative transfer simulations. Using recent fits to numerical relativity simulations, we predict how the ejecta parameters from BNS mergers shape the population of kilonovae, accounting for the viewing-angle dependence. Our simulated population of binary neutron star (BNS) mergers produce peak {\it i}-band absolute magnitudes $-20 \leq M_i \leq -11$. A comparison with detailed radiative transfer calculations indicates that further improvements are needed to accurately reproduce spectral shapes over the full light curve evolution.
\end{abstract}



\section{Introduction}\label{sec: introduction}
Astrophysical understanding of the multi-messenger signals from binary neutron star (BNS) mergers has advanced considerably with the discovery of GW170817/AT2017gfo \citep{TheLIGOScientificCollaboration2017, Abbott2017a, Collaboration2017, coulter_swope_2017, Cowperthwaite2017, Kasliwal2017, lipunov_master_2017, Tanvir2017, Smartt2017, Soares-Santos2017, valenti_discovery_2017}. Extensive analysis of this merger has converged on a description of the optical and infrared (OIR) electromagnetic (EM) signal as being produced by ejected mass with a composition that varies with the polar angle in the frame of the merger \citep{Cowperthwaite2017, Perego2017, Kawaguchi2018}.

The first detection intensified the efforts to realistically model neutron star mergers. While an entirely realistic modelling of all physical aspects is still out of reach, major efforts have been undertaken to explore the parameter space of BNS mergers in terms of masses, mass ratios, equations of state, spins, and other orbital parameters, albeit with approximate physics. With hundreds of available simulations, empirical relations between the intrinsic binary properties and the characteristic kilonova (kN/e) ejecta have been constructed \citep{Kawaguchi2016, Dietrich2016, Coughlin2018, Kruger2020, Nedora2022}.\par

\citet{Dietrich2016}, following the work by \citet{Foucart2012, Kawaguchi2016} for black hole-neutron star systems, provided the first fits of empirical formulae relating the properties of a BNS merger, i.e., the component gravitational masses $M_{1,2}$ and neutron star compactness parameters $C_{1,2}$, to the dynamical ejecta parameters: ejecta mass $m_{\mathrm{ej}}$ and ejecta velocity $v_{\mathrm{ej}}$. The fitting formulae presented in \citet{Dietrich2016} have since been updated in the works by \citet{Coughlin2018, Radice2018, Kruger2020, Nedora2022}. With these relations it has become feasible to investigate the diversity of kNe resulting from a set of priors describing the progenitor BNS population.

Understanding the population of kNe will help resolve systematics in measurements of the Hubble constant from standard sirens \citep{Mortlock2018, Chen2020, Coughlin2020, Moresco2022}, improve population level inference of the nuclear matter equation of state (EOS) from neutron stars \citep{Lackey2014, Wysocki2020, Ghosh2022a}, and improve understanding of the rate of heavy element enrichment of the Universe \citep{Cowan2019}. As the population becomes more understood, observational selection biases can be studied and incorporated into analyses aimed at determining population level parameters, or place informed priors on multi-messenger parameter estimation from such events. A complete population description and resulting observational selection will be ever more important if standard sirens, being independent of distance-ladder calibration uncertainties, are going to be the arbiter of the Hubble-tension \citep{Mortlock2018}.

We present a population model of kNe where the parameters of the progenitor BNS describe the resulting kN transient signal distribution. We extend the kN signal modelling used in our previous work, \citet{Setzer2018}, in several ways. First, we augment the relations presented by \citet{Coughlin2018} in their multi-messenger parameter estimation of GW170817 with expressions from \citet{Radice2018, Radice2018b} for additional components of the ejecta to obtain a total ejecta mass contributing to the kN. Further, using simulation data from \citet{Radice2018}, we derive relations for the mass-averaged composition of the ejecta material as a function of viewing-angle. Then, using a library of detailed nuclear heating rates, we calibrate a semi-analytic grey-opacity kN model with high-fidelity radiation transport simulations; these steps constitute improved physical modelling with respect to recent population models presented by \citet{Nicholl2021} and  \citet{Colombo2022}. However, in contrast to these works, we simplify the computation of our model by neglecting composition differences from off-axis contributions of the ejecta. We then train a Gaussian process emulator to create draws from this calibrated model for a broad parameter space spanning the ejecta properties of the merger. Finally, we study the resulting distribution of kN light curves to get an understanding of the range and diversity of signals that are possible.

This work represents a significant step towards improved modelling for rapidly simulating populations of kNe consistent with BNS merger progenitors and detailed physics, which is necessary to make accurate predictions for detection prospects \citep{Rosswog2016b, Scolnic2017a, Setzer2018, Almualla2021, Mochkovitch2021, Carracedo2020, Andreoni2022, Chase2022, Colombo2022, Just2021}. 

In Sec. \ref{sec:BNS_pop} we detail the choices of the priors on the population of neutron star binaries. Section \ref{sec:kn_model} presents the updates to modelling of the optical-NIR kN signal resulting from BNS mergers. We present the results from these simulations in Sec. \ref{sec: results} and also discuss the dependence of the resulting kN signals on the binary parameters and the impact of modelling uncertainties on the resulting distributions. Finally, we conclude in Sec. \ref{sec: conclusion} with a summary of the implications for future detections of BNS kNe and, discuss the additional astrophysical modelling needed to improve our understanding of these objects.

\section{Population of Kilonova-producing BNS}\label{sec:BNS_pop}
We will now define priors on the set of parameters which will allow us to simulate kNe for each BNS merger. While all mergers of neutron stars are predicted to produce some level of EM emission \citep{Metzger2012}, not all will produce a kN signal that is bright enough to be observed by current and near-term instruments \citep{Setzer2018, Carracedo2020}. This is expected due to the magnitude limit in the context of surveys, and also due to the intrinsic variability in the population. To model the population of BNS mergers in the Universe, we use the broad prior on BNS masses used by the Laser Interferometer Gravitational-wave Observatory and Virgo Collaboration (LVC) template bank-based searches for compact binary coalescence signals \citep{Canton2017}. For our population of BNS, this is a uniform prior on the component masses of the binary from one to three solar masses. However, from this starting point we make three modifications to the component mass prior. 
\begin{figure}
    \centering
    \includegraphics[width=1\columnwidth]{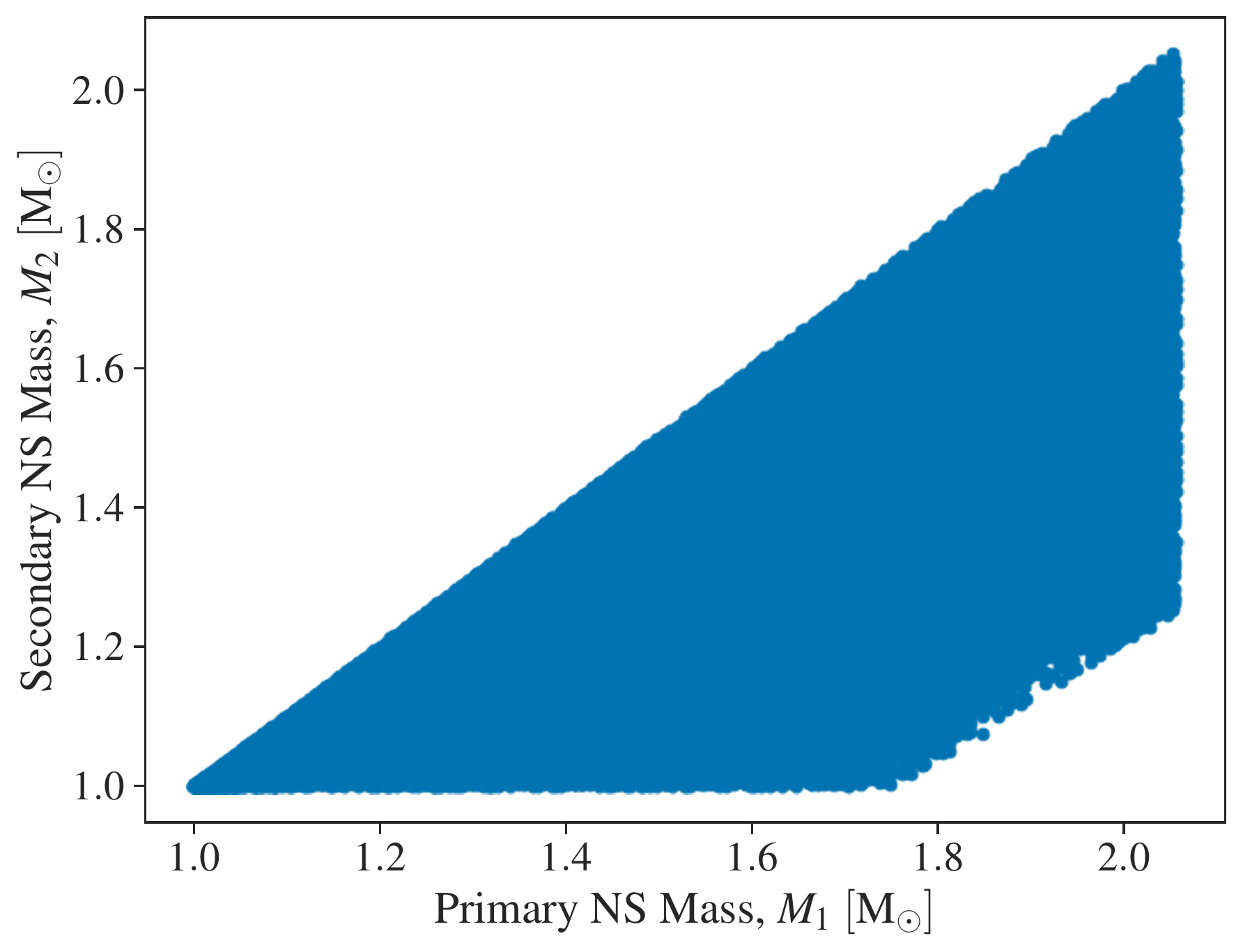}
    \caption{Parameter space of BNS component masses for which we simulate a population of kNe. We begin with the typical range considered for BNS mergers by the LVC searches, modifying it with additional restrictions. The feature in the lower-right corner arises from a limit on the maximum ejecta mass which correlates strongly with mass-ratio as seen in Fig. \ref{Fig:dynamical_ejecta}, rather than the minimum mass-ratio limit. This results in an effective mass-ratio limit of $q\simeq 0.55$.}\label{Fig:binary_mass_prior}
\end{figure}

\begin{table}
\centering
\caption{Prior on the progenitor BNS joint component mass distribution. The starting point is the broad prior used by LVC in their searches for neutron star gravitational wave emission, approximately $1\,-\,3 \ M_{\odot}$. This prior is modified using additional criteria.}
    \begin{tabular}{l|c|c}
        Constraint & Imposed Limits \\ \hline
        Component mass symmetry & $m_2 \leq m_1$ \\
        Maximum mass & $m_1 \leq M_{\mathrm{TOV}}$ \\
        Minimum mass ratio & $q \equiv \frac{m_2}{m_1} \geq \frac{2}{5}$ \\
        Maximum ejecta mass & $m_\mathrm{ej, max} = 0.08$ [$M_\odot$] \\
        \hline
    \end{tabular}
\end{table}
We set the maximum neutron star mass to the maximum Tolman-Oppenheimer-Volkov (TOV) mass given by a choice of equation of state \citep{Tolman1939, Oppenheimer1939}, $M_{\mathrm{TOV}} \approx 2.0-2.5 \, \mathrm{M}_{\odot}$, see \citet{Godzieba2020, Drischler2020} and references therein. We also place a cut on the mass-ratio, $q = m_{\mathrm{min}}/m_{\mathrm{max}}$. We choose to simulate binaries in the range $q \in [0.4, 1]$. This cut is chosen to accommodate the range of mass-ratios inferred from the detections of GW170817 and GW190425 \citep{GW170817props, GW190425props}, while also including the ranges from current observational limits of $q_\mathrm{min, obs} \sim 0.75$ \citep{Martinez2015} and the prediction of mass ratios from stellar population synthesis models of $q_{\mathrm{min}} \sim 0.5-0.6$ \citep{Dominik2013, Tauris2017, Kruckow2018, Andrews2019}. Notably, given an imposed upper limit on the ejecta mass, this effectively limits mass ratios of our population to the range $q \in [0.55, 1]$. The neutron star component masses are drawn uniformly from this joint distribution, see Fig. \ref{Fig:binary_mass_prior}. We note that a portion of our population, where the total remnant mass is less than $1.2 M_\mathrm{TOV}$, will produce a long-lived neutron star remnant and likely have a magnetar-driven kilonovae \citep{margalit_angular-momentum_2022, sarin_diversity_2022}. We do not modify our modelling for this sub-population, though we find approximately 50\% of our population have a predicted total remnant mass below 1.2$M_\mathrm{TOV}$.

Additionally, we assume the binary systems are comprised of non-rotating neutron stars \citep{Bildsten1992, Kochanek1992}. Further assuming that there is no precession of the orbital angular momentum of the binary, we can promote the binary merger-frame polar angle, i.e. the angle of total angular momentum, to the observer viewing-angle, $\theta_{\mathrm{obs}}$. Given this we can equate the observer viewing-angle and the binary orbital inclination. With the further assumption that there is symmetry of the merger ejecta about the binary orbital plane we map viewing-angles greater than $90$ degrees as,
\begin{align}
    \theta_{\mathrm{obs}} = \begin{cases}
\theta_{\mathrm{obs}} \ \ \ \ \ \ \ \ \ \ &\mathrm{for} \ \theta_{\mathrm{obs}} \leq \frac{\pi}{2}, \\
\pi - \theta_{\mathrm{obs}} \ &\mathrm{for} \ \theta_{\mathrm{obs}} > \frac{\pi}{2}.
\end{cases}
\end{align}
Each binary is then oriented assuming no preferential direction in alignment of their inclination. This is realized by drawing the observer viewing-angle/binary inclination angle from a uniform distribution on the sphere. Given we are assuming azimuthal symmetry, i.e., symmetry about the binary orbital axis, this maps to
\begin{equation}
p(\theta_{\mathrm{obs}},\phi_{\mathrm{obs}})=\frac{\sin(\theta_{\mathrm{obs}})}{2 \pi },
\end{equation}
where $\phi_\mathrm{obs}$ is the aziumthal angle of inclination on the unit sphere, which factors out due to axisymmetry.

\subsection{Equation of State}
Assuming all neutron stars obey a single equation of state (EOS), we consider those used by the simulations of \citet{Radice2018}. These EOS are consistent with the tidal deformability constraints from GW170817 \citep{Radice2017}. Though none of the EOS are strongly favoured relative to another; based on the analysis of \citet{Coughlin2018} and due to its general use in the neutron star simulation community, we choose to use the SFHo EOS, with $M_\mathrm{TOV} \approx 2.06 \, \mathrm{M}_{\odot}$ \citep{Steiner2012}. This EOS was designed with consideration towards properties of observed neutron stars and results of near-saturation density nuclear experiments.

The mass-radius relation, combined with the solution of the TOV equations for our given EOS, allows us to calculate the stellar compactness, $C$, given by
\begin{equation}\label{eqn:compactness}
    C = \frac{GM}{c^2 R},
\end{equation}
where $G$ is the gravitational constant, $c$ is the speed of light, and $R$ is the neutron star radius. Given the range of variation allowable from the choice of EOS we will consider only EOS without a neutron star crust model. With this specification of EOS we now have completed our priors on the kN-progenitor binary systems. We will now describe how these parameters are mapped to the inputs necessary for simulating the counterpart kN signal.

\section{Modelling of BNS Kilonovae}\label{sec:kn_model}
We now describe the process of modelling each kN, beginning with a brief summary of the ejecta components which contribute to the kN-emission. To model a kN dependent on the BNS parameters described in Sec. \ref{sec:BNS_pop}, we map those parameters to those of the kN model. Our kN model characterizes the ejecta with four parameters: the total amount of ejected matter, $m_\mathrm{ej,\, total}$, the median outflow velocity, $v\ej$, the electron fraction of the material, $Y_\mathrm{e}$, and the grey opacity of the ejecta, $\kgrey$. The first three parameters are determined via empirical mappings from numerical simulations, discussed in Sec. \ref{sec:empirical_mappings}. In Sec. \ref{sec:nuc_mat_rels}, we describe the nuclear heating rate prescription we use and the radiation transport simulations used for calibrating grey-body opacities, which we compute with {\sc SuperNu} \citep{Wollaeger2013, Wollaeger2017, Even2019a}. The process for determining the final parameter, the grey opacity, is described in \ref{sec:grey_opac}. The process of calibrating the grey-opacities of the kN model with the additional radiation transport simulations is described in Sec. \ref{sec:greyfit}. In Sec. \ref{sec:GP_emul}, we use these results to train a Gaussian process emulator to predict grey-opacities over the entire ejecta parameter space.

\subsection{Phenomenological Description}\label{sec:pheno_desc}
As the inspiral of a binary neutron star system progresses, the two stars approach each other and will eventually come under the influence of the tidal forces of their companion beginning the process of ejecting material from the system; then, in their collision, squeezed and shock-heated material will also be ejected from the gravitationally bound system \citep{oechslin_relativistic_2007, Rosswog2015, Tanaka2016, Metzger2019}. These processes happen on the dynamical time-scale of the merger and form what is called the dynamical ejecta. Additional components of the ejecta occur from neutrino winds, magneto-rotational instabilities in the accretion disc, and secular processes that occur during the formation of the post-merger remnant, i.e. a super-massive or hyper-massive neutron star or a stellar mass black hole \citep{Rosswog2015, Radice2018, Radice2018, Metzger2019, Sarin2020}. Assuming that the BNS inspiral is circular, and the binary is comprised of two non-spinning neutron stars, the resulting ejecta is expected to be symmetric about the merger plane. However, along the polar angle in the merger-frame, the properties of the ejecta will vary. Related back to the observer, this creates a viewing-angle dependence of kNe.

While each of these components of the ejecta have their own properties, and likely different composition profiles, we make the assumption that the light curves will be dominated by the contribution from the outermost ejecta in the observer's line-of-sight. This simplifying assumption allows for significant advantages in terms of decreased model complexity and computational efficiency. Further, \citet{Kawaguchi2019} indicate the outer dynamical ejecta do occlude, at least partially, contributions from the other components, suggesting their effect on the light curves is secondary. Given these assumptions we can then model the kN, including viewing-angle dependencies, by solving a 1D homologous flow radiation transport model with inputs specified by the total ejecta mass, median ejecta velocity, and the line-of-sight composition of the outermost material. 

\subsection{Connecting Ejecta Parameters to BNS Parameters}\label{sec:empirical_mappings}
We now describe how we determine the bulk ejecta properties and line-of-sight composition, which are inputs into our model. Sampling the parameters $\{M_1,M_2,\theta_{\mathrm{obs}}\}$ from the population prior, with a fixed EOS, we construct a mapping to all other parameters describing the simulated kN-signals $\{m\ej, v\ej, Y_\mathrm{e}, \kgrey\}$. 

\subsubsection{Mapping Binary Parameters to Kilonova Ejecta Properties}\label{sec:self_consistency}
Fully self-consistent simulations of BNS mergers from inspiral, General Relativistic (GR) merger, to GR-Magneto-Hydrodynamic (GR-MHD) outflow with fully relativistic radiation transport are immensely computationally expensive, even for a single event. We approach modelling of the BNS merger population using empirical relations derived from a substantial number of merger simulations which relate the binary parameters to the parameters of the ejecta producing the kN signal. We adopt the relations from \citet{Coughlin2018} which present updates to the fitting formula of \citet{Radice2018} for $m_{\mathrm{ej}}$ and $v_{\mathrm{ej}}$ using 259 numerical simulations. They have found decreased error by fitting $\log_{10} (m_{\mathrm{ej}})$ instead of $m_{\mathrm{ej}}$ and have simplified the relations by removing the need for solving for the component baryonic masses \citep{Coughlin2018}. The equations and parameters are
\begin{align}
\log_{10} (m_{\mathrm{ej}}^{\mathrm{fit}}) &= \left[a \frac{(1-2C_1)M_1}{C_1} + b M_2 \left(\frac{M_1}{M_2}\right)^n + \frac{d}{2} \right] + [1 \leftrightarrow 2], \label{eqn:mej_mapping} \\
v_{\mathrm{ej}}^{\mathrm{fit}} &= \left[e (1+fC_1)\left(\frac{M_1}{M_2}\right) + \frac{g}{2} \right] + [1 \leftrightarrow 2], \label{eqn:vej_mapping}
\end{align}
with $a=-0.0719,\, b=0.2116,\, d=-2.42,\, n=-2.905,\, e=-0.3090,\, f=-1.879,\, \mathrm{and\,} g=0.657$, where $[1 \leftrightarrow 2]$ refers to repetition of the preceding fitting terms with the binary parameter indices interchanged, i.e. $M_1 \leftrightarrow M_2$ \citep{Coughlin2018}. This corresponds to a fractional error of 36\% in $\mathrm{log}_{10} (m_{\mathrm{ej}}^{\mathrm{fit}})$, equation \ref{eqn:mej_mapping} (dominated by the low-mass end, $m_{\mathrm{ej}}\approx 10^{-4}-10^{-5}$, though significantly better for larger $m_{\mathrm{ej}}$) and 18\% in $v_{\mathrm{ej}}$ \citep{Coughlin2018}.

To obtain the ejected matter contributions to the total ejecta, we include additional relations for the secular ejecta coming from \citet{Radice2018} and \citet{Coughlin2018}. These primarily include an estimate for the remnant disc mass, $m_\mathrm{disc}$ given by:
\begin{align}
    \mathrm{log}_{10}(m_\mathrm{disc}&[M_\mathrm{tot}/M_\mathrm{thresh}]) = \nonumber \\ &\mathrm{max}\left(-3, a\left(1+b \, \mathrm{tanh}\left[\frac{c - M_\mathrm{tot}/M_\mathrm{thresh}}{d}\right]\right)\right),
\end{align}
where $M_\mathrm{tot}$ is the total mass of the merging binary, $M_\mathrm{thresh}$, is the mass threshold for prompt black hole collapse, and $a=-31.335,\, b=-0.9760, \, c=1.0474,\, d=0.05957$ are the fitting parameters to numerical data. The black hole prompt collapse threshold mass is computed following \citet{Bauswein2013}, which incorporates the chosen EOS through specification of the TOV maximum mass, $M_\mathrm{TOV}$, and the neutron star radius at 1.6 solar masses, $R_{1.6M_\odot}$,
\begin{equation}
    M_\mathrm{thresh} = \left(2.38 - 3.606 \frac{M_\mathrm{TOV}}{R_{1.6M_\odot}} \right)M_\mathrm{TOV}.
\end{equation}
This contributes to the total ejecta given an efficiency of unbinding of the disc \citep{Radice2018, Radice2018b}, $\eta_\mathrm{disc}$, due to secular processes, i.e., neutrino winds, GRMHD instabilities, etc.
\begin{equation}
    m_\mathrm{secular} = \eta_\mathrm{disc} m_\mathrm{disc}.
\end{equation}
Given numerical results which show that the unbinding of the disc can contribute between $10\% - 40\%$ of the total disc mass to the ejecta, we sample uniformly between these two bounds to obtain the percentage of unbound material contributing to the ejecta for each kNe powering the EM transient \citep{Metzger2008, Siegel2017, Miller2019a, Fernandez2018a}. Thus the total ejecta mass for each kN is given by,
\begin{equation}
    m_\mathrm{ej,\, total} = m_\mathrm{ej}^\mathrm{fit} + m_\mathrm{secular}.
\end{equation}

\begin{figure}
    \centering
    \includegraphics[width=1\columnwidth]{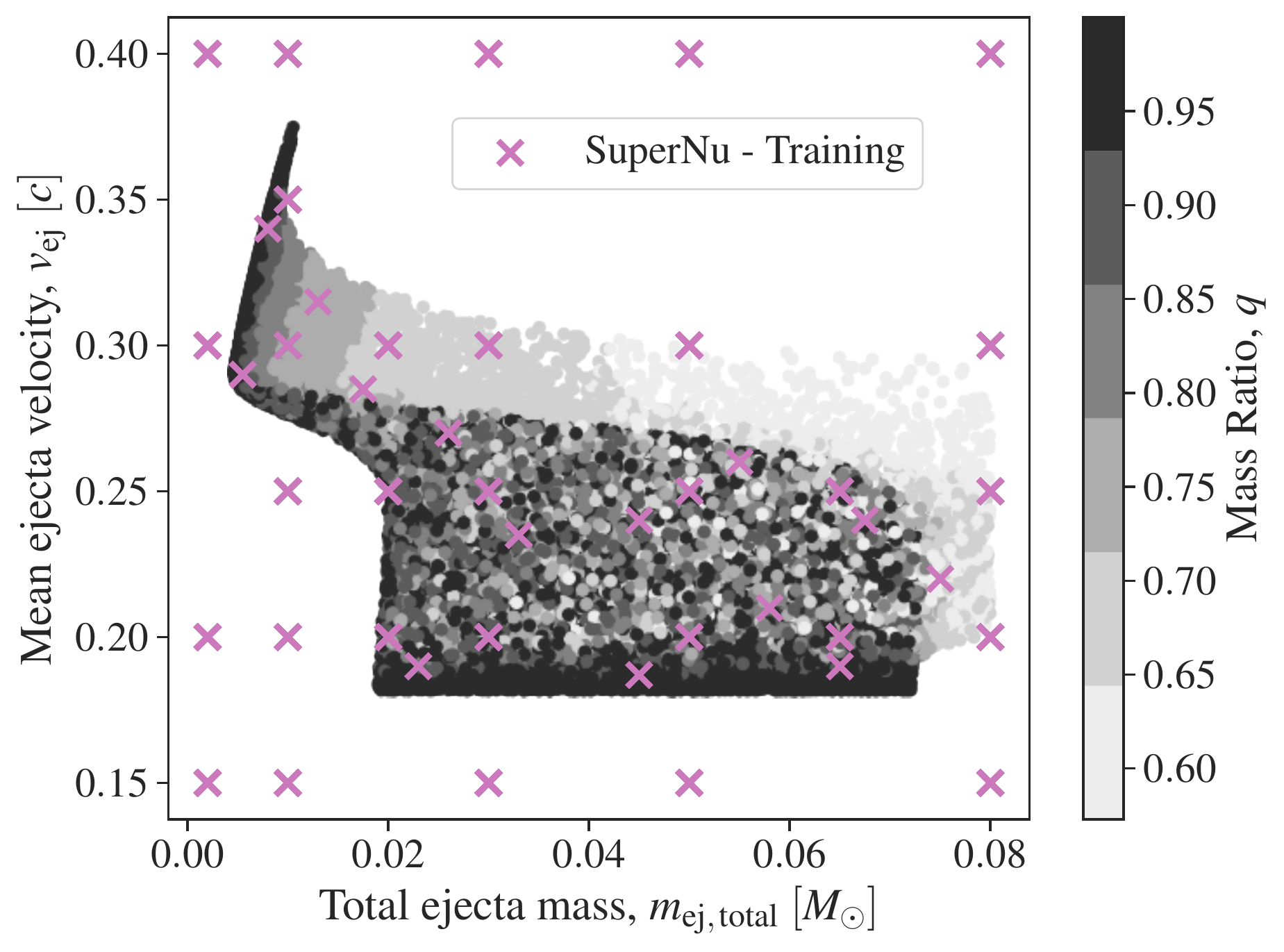}
    \caption{Ejecta parameter space given by the prior on the binary masses, Fig. \ref{Fig:binary_mass_prior}, mapping equations \ref{eqn:mej_mapping} - \ref{eqn:vej_mapping}, and the SFHo EOS. The points are shaded by the mass ratio of the neutron stars comprising the binary which maps to each ejecta parameter pairing. Over-plotted are the locations in the ejecta parameter space of the training data used to calibrate the grey opacity of the model to {\sc SuperNu} radiation transport simulations. While this is a complex shape given the non-linear mappings of the parameters, we can identify the swooping feature at the left of the figure to arise from merger scenarios where the remnant object undergoes prompt collapse to a black hole, thus capturing more material and preventing efficient formation of an accretion disc from which additional material can be ejected through MHD-turbulence and other secular processes.}\label{Fig:dynamical_ejecta}
\end{figure}

\begin{table}
\centering
\caption{Approximate regions of allowable ejecta parameter space. This is meant to parameterize, crudely, the "elephant"-shaped region of ejecta parameter combinations allowable from the prior and mappings used in this work shown in Fig. \ref{Fig:dynamical_ejecta}.}
    \begin{tabular}{c|c|c|c}\label{Tab:apprx_ej_params}
         Region & Coordinates of Approximate Boundary Vertices  \\
          & ($m_\mathrm{ej}$, $v_\mathrm{ej}$)   \\ \hline
        1 & (0.005, 0.275) \\
        &  (0.01, 0.375) \\
        &  (0.01, 0.34) \\
        &  (0.02, 0.32) \\
        &  (0.02, 0.175) \\ \hline
        2 & (0.02, 0.175) \\
        & (0.02, 0.32) \\
        & (0.08, 0.3) \\
        & (0.08, 0.175) \\
        \hline
    \end{tabular}
\end{table}

The ejecta parameters resulting from these mappings are shown in Fig. \ref{Fig:dynamical_ejecta} and produce a complex shape with a significant concentration of kNe at high velocity and low ejecta mass. As noted by \citet{Radice2018}, these fitting formulae do not capture all the effects that contribute to the ejecta. Additionally, the numerical relativity simulations which comprise the basis for these fits do not all simulate significantly beyond the time of merger. Further, the detailed microphysics of neutrino transport, magnetohydrodynamic (MHD) turbulence, and viscous effects are not usually simulated simultaneously. However, given the lack of data for BNS mergers and remaining uncertainty in the physics of these mergers, the above relations are sufficiently robust for simulating signals from a cosmological population of BNS mergers. Further investigating results from numerical simulations will allow us to encode additional dependencies of BNS kN ejecta on properties of the binary merger, such as spins and tidal deformability, which we leave to future work.
\subsubsection{Viewing-Angle Dependence}\label{sec:observer_angle}
Using the data from the simulations of \citet{Radice2018}, we adopt the description of the dynamical kN-ejecta properties as a function of the binary merger-frame polar angle $\theta$ from \citet{Perego2017}; see fig. 2 from their work. This uses the electron fraction, $Y_\mathrm{e}$, as the primary property describing the composition of the material. We sub-select simulations from this dataset which do not promptly form a black hole \citep{Agathos2019, Kolsch2021}, i.e., $t_{\mathrm{BH}}\leq 2\, \mathrm{ms}$ as this would not expected to be accompanied by an observable EM-counterpart due to extremely low total ejecta mass \citep{Margalit2017}. These simulation outputs are divided into angular slices which each contain a distribution of matter with different properties, e.g., fig. 4 of \citet{Radice2018}.

\begin{figure}
    \centering
    \includegraphics[width=1\columnwidth]{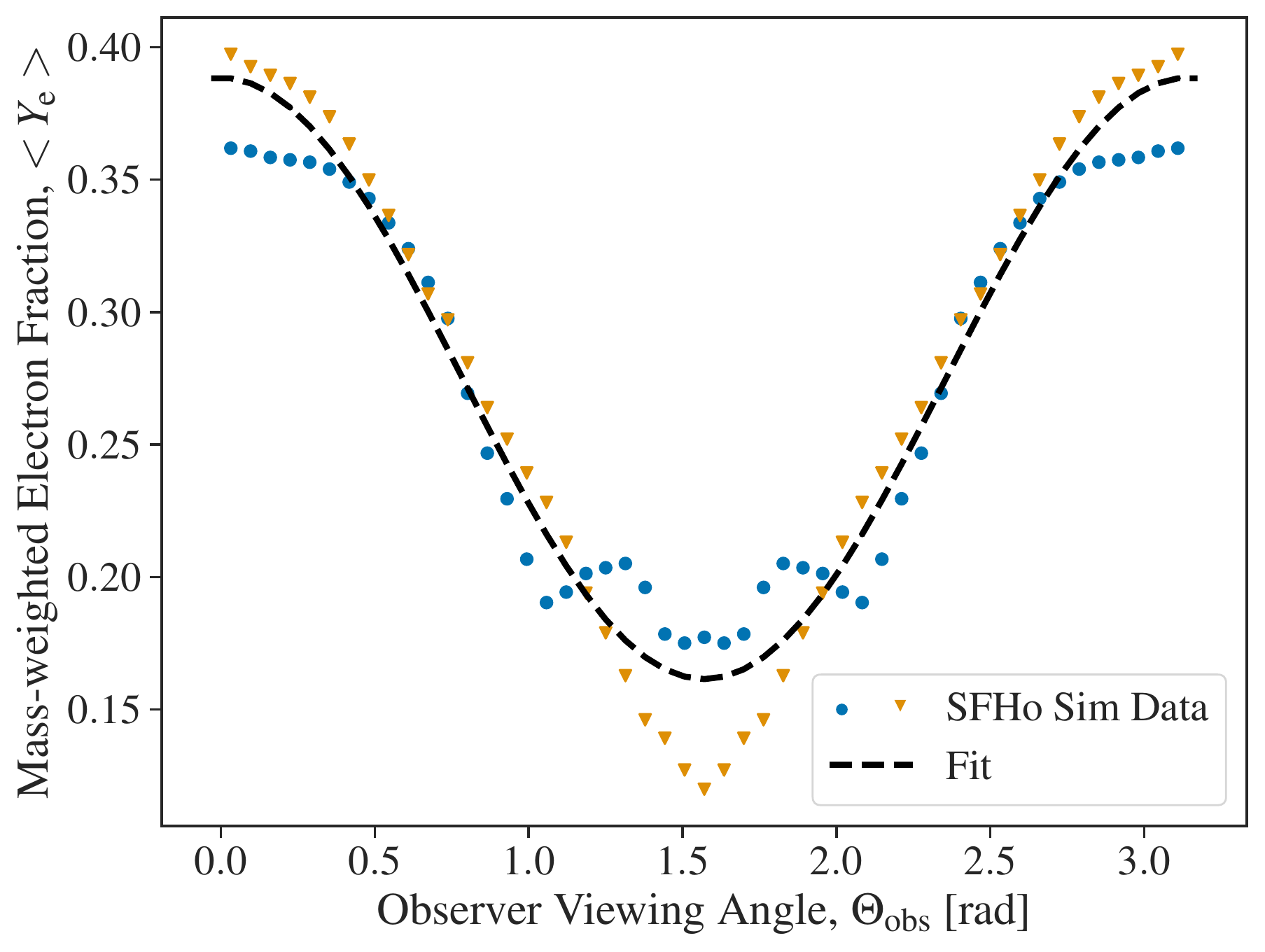}
    \caption{Fit for the mass-weighted average electron fraction, $\langle Y_{\mathrm{e}} \rangle (\theta)$ for the SFHo equation of state with simulation data coming from a subset of \citet{Radice2018} which have a more detailed prescription for the microphysics. Deviations from the fit are greatest around the merger plane, i.e., a viewing angle of $\pi /2$, where the structure of the electron fraction distribution becomes more complex. The two sets of points in the figure represent simulations with different initial conditions for the constituent masses of the binary. The blue circles correspond an equal-mass neutron star binary of two $1.35 \ M_\odot$ neutron stars and the orange triangles correspond to a $1.4 \ M_\odot$ -- $1.2 \ M_\odot$ binary neutron star system.}\label{Fig:SFHo_ye_fit}
\end{figure}

We take the mass-weighted average of the electron fraction data along each angular slice to obtain a profile of $\langle Y_{\mathrm{e}}\rangle$ vs. $\theta$. Noting the findings of \citet{Perego2017}, a simple trigonometric function is able to describe the ejecta mass as a function of polar angle, i.e., $\mathrm{m_{ej}}(\theta)\approx \sin^{2}(\theta)$. We obtain a similar description of the mass-weighted electron fraction, $\langle Y_{\mathrm{e}} \rangle$, by fitting trigonometric functions to the data.
The best fit function for those tested, by least-squares optimisation, is for
\begin{equation}\label{eqn:Ye_fit_func}
\langle Y_{\mathrm{e}} \rangle(\theta) = a \cos^2(\theta) + b,
\end{equation}
where $a = 0.22704$ and $b=0.16147$ are the fit parameters for the SFHo EOS. We improve this fit further by splitting the simulation data based on the underlying equation of state and also selecting the subset of physics which includes not only neutrino cooling, but also neutrino heating. This reduces fit deviations substantially to $\mathrm{max}(\Delta\langle Y_{\mathrm{e}}\rangle) \approx 0.05$, see Fig. \ref{Fig:SFHo_ye_fit}. We note that including additional parameters, i.e., the total binary mass and the binary mass ratio, at linear order did not show measurable improvement. The electron fraction composition is the primary determining factor of the nuclear heating rates and grey opacity which is described next.

\subsection{Nuclear Heating}
\label{sec:nuc_mat_rels}
The nuclear heating rate, $\dot\varepsilon(t)$, in the ejecta consists of the energy released during decays of a large number of radioactive isotopes which are produced in the rapid neutron-capture process ($r$-process). Although it has been demonstrated that such heating can be well approximated by the power law $\dot\varepsilon(t)\sim t^{-\alpha}$ \citep{Metzger2010, Hotokezaka2017_plaw}, with $\alpha\approx1.2-1.3$, the accuracy of this approximation is insufficient for our purposes. We therefore created a library of nuclear heating rates, parameterized by the initial electron fraction $Y_{\rm e}\in [0.05,0.5]$ and ejecta velocity $v_{\rm ej}\in[0.05c, 0.4c]$. 
The nuclear heating rate was computed on a grid of these parameters using a nucleosynthesis network {\sc WinNET} \citep{Winteler2012, Korobkin2012}, an update of the {\sc BasNet} network \citep{Thielemann2011}. This is the same nucleosynthesis code which was used to compute the $r$-process nucleosynthesis in \cite{Wollaeger2017}, using 5831 isotopes and the reaction rates from the compilation of \cite{Rauscher2000}. The nuclear masses far from stability are not experimentally known and one has to resort to theoretical mass models. In our network, we use the finite range droplet model \citep[FRDM;][]{Moeller1995}. The weak interaction rates are taken from \cite{Arcones2011}. For fission and neutron capture, the fission rates of \cite{Panov2010} and $\beta$-delayed fission probabilities of \cite{Panov2005} were used.

The nucleosynthesis calculations are performed along ejecta trajectories whose density as a function of time is computed from the ejecta mass $m_{\rm ej} = 0.05\ M_\odot$ and expansion velocity $v_{\rm ej}$ on the grid. The initial temperature is computed from the equation of state using the initial density and entropy, which is adopted to be $s_{\rm ej} = 15\ k_B/{\rm baryon}$. Later, during the evolution of abundances, the entropy is incremented according to the produced heat, and the temperature is computed accordingly following \citet{1999ApJ...525L.121F}.

For each of the 120 points on the $\{Y_{\rm e}, v_{\rm ej}\}$-grid, the nuclear heating rate is fit with an approximate formula, that has 11 fitting coefficients. The latter are then interpolated to obtain the values of fitting coefficients for the ejecta parameters in between the values of the grid. We describe the fitting procedure and the nuclear heating rates library elsewhere \citep{rosswog_heavy_2022}. Both the full blown radiative transfer simulations with {\sc SuperNu} and our simpler semi-analytic model use the same nuclear heating rates from this library for the kN light curves calculations.

\subsubsection{Density-averaged time-dependent thermalisation}
We additionally implement time- and density-dependent thermalisation efficiencies following \citet{barnes_radioactivity_2016}, and \citet{Wollaeger2017}, and recently summarized in sec. 2.2 of \citet{Bulla2022}. Instead of computing the efficiencies over the entire density profile of the ejecta, we replace the density profile $\rho(t, \mathbf{r})$ with an averaged density. The details of our implementation are given in App. \ref{sec:appendix}.

\subsection{Grey Opacity Dependence on Ejecta Properties}\label{sec:grey_opac}
In order to produce a computationally inexpensive model useful for parameter estimation and rapid interpretation of observations, we employ a grey-opacity model for the kN emission. In general, the opacity of a kN has contributions from a large number of lines due to the presence of heavy elements \citep{Kasen2013}. The number of these lines can be greater than $10^6$ for ions of some lanthanides and actinides, such as Terbium, Erbium, or Protactinium. However, we intend to summarize this with a single grey-opacity. As this is a significant simplification of the physics, we expect the grey opacity we infer to have dependencies on other parameters of the kN, such as the density and expansion-rate of the material. Thus we approach the mapping to grey-opacity including all parameters of the ejecta derived to this point, i.e., $\{m\ej, v\ej, Y_\mathrm{e}\}$.
\begin{figure*}
    \hspace{-1.0cm}
    \centering
    \includegraphics[width=\textwidth]{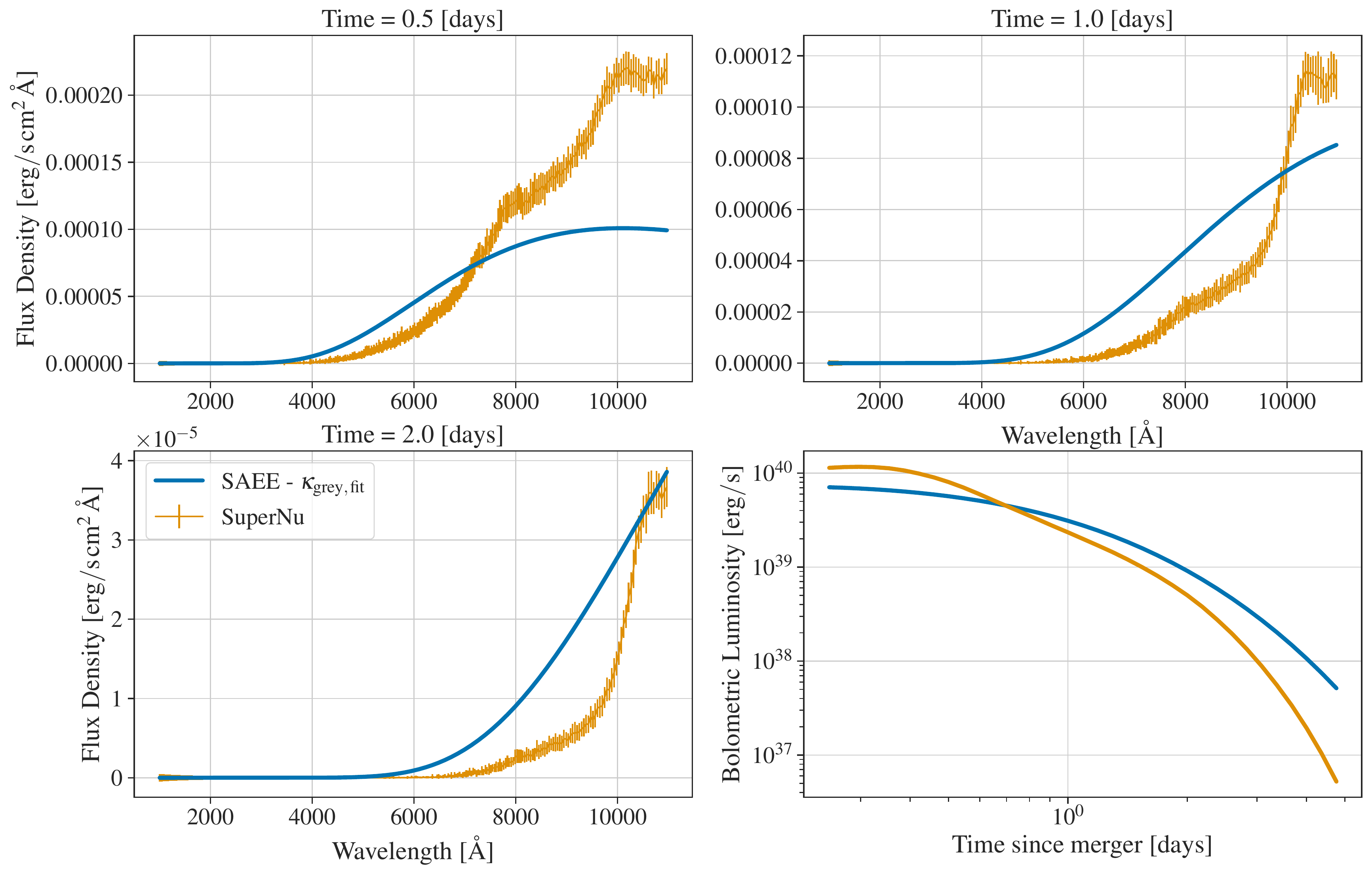}
    \vspace{-0.15cm}
    \caption{Snapshots of the SED timeseries and overall bolometric luminosity evolution for a single model from our {\sc SuperNu} training set data and the grey-body model fit. The ejecta parameter values for this simulation are $m\ej=0.03\, [M_{\odot}],\, v\ej=0.25 \, [c],\, Y_\mathrm{e}=0.15$. This model represents a typical event drawn from the population. The upper-left panel shows the spectrum closest to peak luminosity of the grey-body model. The other epochs shown are typical of the time scales at which it is expected observational followup of kNe to be made. We see good agreement in the relative color, i.e. shape, of the spectrum, although the model tends to systematically under-predict the overall luminosity, i.e., the amplitude. The agreement of the bolometric luminosity, as seen in the lower-right panel, is good over the range of evolution most likely detectable, i.e., less than 2 days, with deviations increasing at later times. Note the grey-body fit uses the same ejecta parameters, i.e., $m\ej,\, v\ej,\, Y_\mathrm{e}$, as  those used in the {\sc SuperNu} simulation and only $\kappa_{\mathrm{grey}}$ is solved for, as described in Sec. \ref{sec:greyfit}.}\label{Fig:model_mistmatch}
\end{figure*}
\subsubsection{Grey-body Model and Opacity Fitting Scheme}\label{sec:greyfit}
To accomplish this we compute synthetic spectra from multi-group radiative transfer simulations using {\sc SuperNu} \citep{Wollaeger2013, Wollaeger2017, Even2019a}, for a range of kN-like ejecta trajectories. We choose a grid of kN-parameters, $\{m_{\mathrm{ejecta}},\ v_{\mathrm{ejecta}},\ Y_{\mathrm{e}}\}$, that spans the range of ejecta values predicted for the population of BNS kNe we are simulating to produce our training set. This represents a grid of the following parameter values: $\Theta^\mathrm{train}\ej = \{m\ej \in (0.002, 0.01, 0.03, 0.05, 0.08),\, v\ej \in (0.05, 0.10, 0.15, 0.20),\, Y_\mathrm{e} \in (0.10, 0.15, 0.20, 0.25, 0.30, 0.35, 0.40)\}$. However, to reduce the number simulations necessary to create the training set, we remove a subset of the grid that falls significantly outside the region of the population's parameters as shown in Fig. \ref{Fig:dynamical_ejecta}. This is also augmented with 14 additional simulations offset from the above grid, resulting in a total of 230 simulations in our training set. The results of these {\sc SuperNu} simulations provide
\begin{align}
&\hat{f} = \hat{f}_{i,j}(\bm{\theta}_\mathrm{ejecta}), \\
&\hat{\sigma}^2 = \hat{\sigma}^2_{i,j}(\bm{\theta}_\mathrm{ejecta}), \\
&\mathrm{where:}\ \bm{\theta}_{\mathrm{ejecta}} = \{m_{\mathrm{ejecta}},\ v_{\mathrm{ejecta}},\ Y_{\mathrm{e}}\},
\end{align}
$\hat{f}$ is the spectral flux density, $\hat{\sigma}^2$ is the variance of the spectral flux density, $i$ is the index for the wavelength bins running from $1,..., N_{\lambda}$ and $j$ indexes the time-steps of the simulation from $1,...,N_t$. Note, we also allow the number of wavelength bins to be dependent on the time-step, i.e., $N_{\lambda, j}$. While simulations directly record the total flux per wavelength bin, $\hat{F}$, this is converted to spectral flux density, $\hat{f}$, for the output by dividing the flux by the width of the corresponding wavelength bin, i.e.,
\begin{equation}\label{eqn:flux_to_density}
\hat{F}_{i,j} = (\lambda_{\mathrm{bin},q+1} - \lambda_{\mathrm{bin},q})\hat{f}_{i,j}.
\end{equation}
We fit the {\sc SuperNu} spectral time-series with our grey-body model. We model the kN signal from each BNS merger using a semi-analytic eigenmode expansion (SAEE) model presented by \citet{Wollaeger2017, Rosswog2018} and previously used in \citet{Setzer2018}, see appendix A of \citet{Rosswog2018} for a comprehensive summary of the radiation transport physics. For reference, the spectral time-series is given by a blackbody with effective temperature that evolves according to eq. A.25 of \citet{Rosswog2018}. We make several modifications to the underlying model which solves the diffusion equation and the spectral generation scheme. We introduce an additional parameter, the electron fraction, to model the viewing-angle dependence and to determine the heating rates, see Sec. \ref{sec:observer_angle} and Sec. \ref{sec:nuc_mat_rels}. This model, similarly to {\sc SuperNu}, is parameterized by the ejecta mass, ejecta velocity, and the electron fraction of the ejecta. However, it also contains an additional parameter, the grey opacity, $\kappa_{\mathrm{grey}}$, i.e. the spectral flux density time-series produced by the model is a function
\begin{equation}
s = s_{i,j}(\kappa_{\mathrm{grey}},\, \bm{\theta}_\mathrm{ejecta}).
\end{equation} \par
We fit the grey opacity of the model to the spectral data from {\sc SuperNu} using a weighted chi-sq. method, see Fig. \ref{Fig:model_mistmatch} for a representative example of this fit.\footnote{For reference the grey-body model can be evaluated in approximately 0.08 CPU seconds as compared to 2 CPU hours for {\sc SuperNu}.}

\subsubsection{Restriction to Observationally Relevant Data}
We are concerned with fitting the kN-data most accurately near peak luminosity, as this is when the transient would most likely be detected, enabling follow-up observations to be triggered. Additionally, we are concerned with emulating the signals as they would be observed by optical and near-infrared surveys. For these reasons, we make the following modifications to the data:
\begin{itemize}
    \item We remove {\sc SuperNu} data prior to 0.25 days, as this time-period is undergoing numerical relaxation from the initial conditions to a stable evolution.
    \item We remove {\sc SuperNu} data in the far infrared, keeping only wavelength bins with $\lambda < 11000 \, [\mathring{A}]$.
    \item To prioritize detectability of the kN, we weight the time period closest to peak luminosity higher with respect to the contributions to the total chi-sq. per model, see below.
    \item We remove data after 5 days, as this time-period is beyond the range when LTE radiation transport is reliable for kNe \citep{Pognan2021}.
\end{itemize}
The weighting scheme we adopt places a weight on a given time-step based on the relative luminosity of that time-step with respect to the peak luminosity of the {\sc SuperNu} model being fit. The weights are
\begin{equation}
    w_j = \frac{\hat{L}_j}{\mathrm{max}(\hat{L}_{j})},
\end{equation}
where the luminosity of each time-step of the {\sc SuperNu} data, $L_j$, is defined from the provided data as
\begin{equation}
 \hat{L} = \int_{\lambda_{\mathrm{min}}}^{\lambda_{\mathrm{max}}} \hat{f} \, \md \lambda.
\end{equation}
This weighting is implemented as a re-scaling of the errors that enter the chi-sq. calculation. Explicitly the re-scaled flux density errors, $\hat{\sigma}_{i,j}$, are given by
\begin{equation}
    \hat{\sigma}_{i,j} \rightarrow \frac{\hat{\sigma}_{i,j}}{w_j},
\end{equation}
which then modifies the chi-sq. in the following manner,
\begin{equation}\label{eq:weighted_chisq}
\chi^2 = \sum_j^{N_{t}} w_j^2 \sum_i^{N_{\lambda,j}} \frac{(\hat{f}_{i,j}-s_{i,j} )^2}{\hat{\sigma}^2_{i,j}}.
\end{equation}
Having explored several functional forms for the weighting, we adopt the above due to its simplicity and the fact that it appears to saturate how well the temperature evolution of the {\sc SuperNu} spectra can be fit with a simplified grey-body model. 

For each set of spectra, we fix $\bm{\theta}_{\mathrm{ejecta}}$ for the SAEE model to those from the simulation, and find the value of grey opacity which minimizes equation \ref{eq:weighted_chisq}. Finding the corresponding $\kappa_{\mathrm{grey}}$ for all simulated points $\bm{\theta}_{\mathrm{ejecta}}$ we define a sparse mapping to grey-opacity, i.e., $\kappa_{\mathrm{grey}}(\bm{\theta}_{\mathrm{ejecta}})$. This approximates a grey opacity surface spanning the ejecta parameters of interest. See Fig. \ref{Fig:model_mistmatch} for an example fit from this procedure. \par

\begin{figure}
    \centering
    \includegraphics[width=1.0\columnwidth]{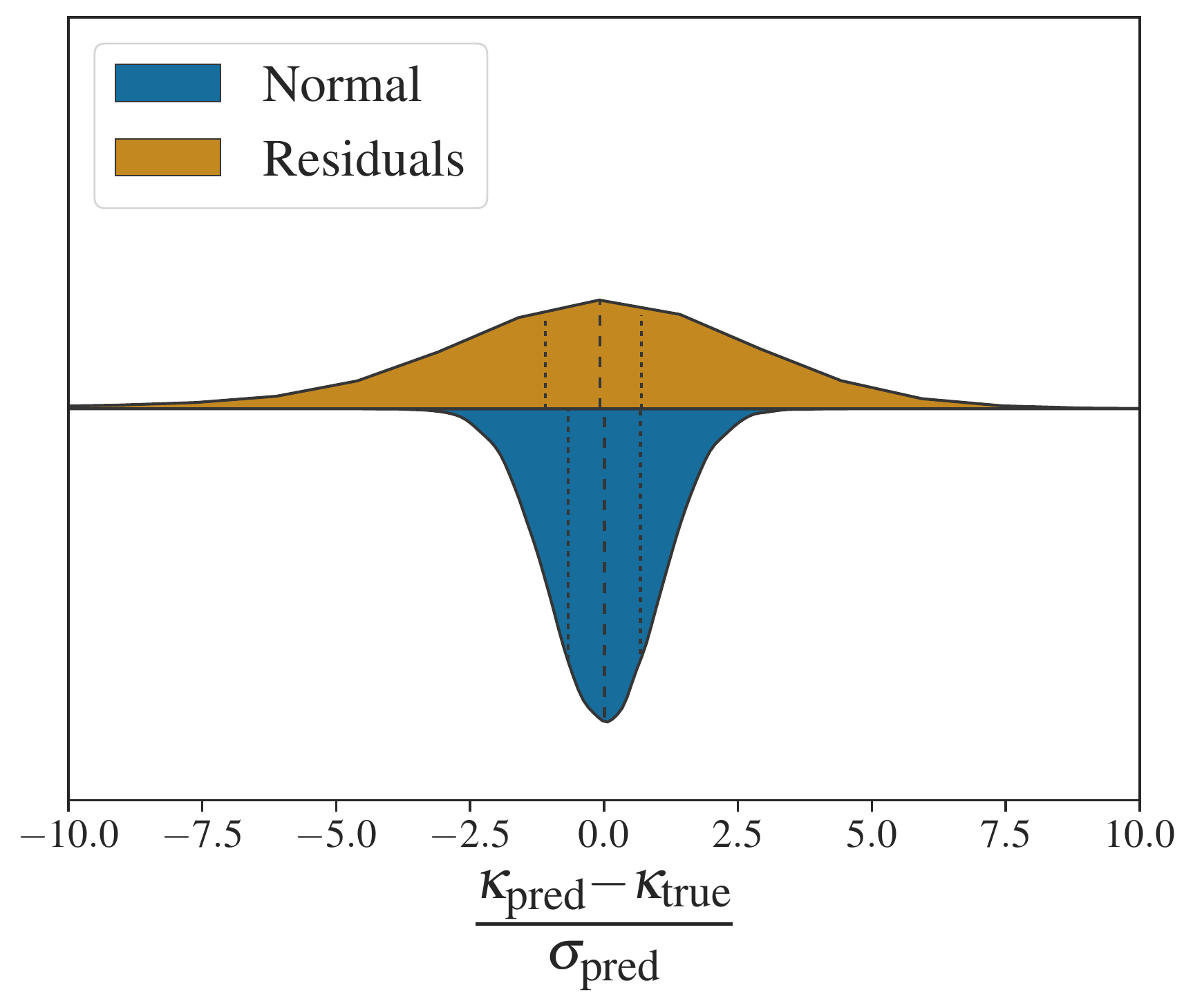}
    \caption{Leave-one-out cross-validation analysis of our chosen kernel, {\it Matern 5/2}. We show standardized residuals from a leave-one-out posterior predictive test of the emulator in comparison to a unit Gaussian distribution with zero mean. While there is a mild skew towards under-predicting the opacity, we note the large outliers are few and none are biased across the grey opacity threshold commonly used to approximate detectability, i.e., $\kgrey \approx 10 \, \textrm{cm}^2 / \textrm{g}$.
    }\label{Fig:GP_LOO}
\end{figure}
\subsubsection{Gaussian Process Emulation}\label{sec:GP_emul}
To extend this mapping to arbitrary parameter combinations we interpolate this surface using Gaussian processes with the package {\sc george} \citep{Ambikasaran2016}. We train on the set of 230 simulations represented by (x)-marks in Fig. \ref{Fig:dynamical_ejecta}. Two simulations were removed from this set due to clear fit failures and outlier chi-sq. values. Given the observed variation of the inferred values on the ejecta properties, we model the covariance of each dimension independently for any chosen kernel function. After exploring a handful of standard kernels, we have chosen to use the {\it Matern 5/2} as the final implementation for our results \citep{Genton2001, Rasmussen2006}. This was chosen over the other kernels as it produced a solution which largely avoids the non-physical region of negative opacity values. \par

This kernel as implemented in {\sc george}, is given by \citep{Ambikasaran2016}:
\begin{align}
    &k(r^2) = \left(1 + \sqrt{5r^2} + \frac{5 r^2}{3} \right)e^{-\sqrt{5r^2}}, \\
    &\mathrm{where} \, r^2 = (\mathbf{x}_i - \mathbf{x}_j)^T C^{-1}(\mathbf{x}_i - \mathbf{x}_j). \nonumber
\end{align}
Here $r^2$ is the squared distance, given the metric $C$, and $\mathbf{x}_{}$ represent the input data coordinates with row, column indices $i,j$. The matrix elements of $C$, and an overall amplitude of the kernel, are the hyperparameters that are optimized to minimize the log-loss of the Gaussian process with respect to the specified mean function given the training data above, $\Theta^\mathrm{train}\ej$. In our scenario we have set the off-diagonal terms to zero and optimized the diagonal terms independently. We construct a piece-wise mean function based on the results of \citet{Tanaka2019}.

We test this interpolation scheme by performing a leave-one-out cross-validation test, see Fig. \ref{Fig:GP_LOO}. This assesses the predictive quality of the emulator by training the emulator on the original training data leaving out one data point at a a time, predicting the value of that held-out datum, and iterating in this manner through the entire training set. In each iteration, the residual between the prediction at the location of the removed data point and the ground truth value from the training data, divided by the predicted emulator uncertainty at that point, is computed. This procedure yields a distribution of residuals, as shown in Fig. \ref{Fig:GP_LOO}. For a perfect emulator, this distribution of residuals should match a Gaussian distribution with unit standard deviation and zero mean (shown for comparison in the figure). We find that the emulator mildly skews towards under-predicting the opacity, i.e., more negative residuals; however, this bias occurs in the extremely lanthanide-rich region where opacities inferred from the training data can reach more than $50 \ \mathrm{cm}^{2} \, \mathrm{g}^{-1}.$ The bias does not cause any of the predicted opacities to cross the approximate detectability threshold of $10 \ \mathrm{cm}^{2} \, \mathrm{g}^{-1}$ \citep{Setzer2018}. Consequently, we do not expect this to significantly impact predictions for observations, though the presence of this bias does reflect the possibility for future improvements to the Gaussian process construction for this emulation.

In order to use this in forward-modelling of BNS merger kNe we use the trained Gaussian process to predict the opacity given the arbitrary ejecta parameter combinations of each simulated source. The Gaussian process predicts a mean value and an uncertainty; thus, we sample from this distribution to obtain an opacity for each simulated kN. To avoid unphysical opacities, we reject values below $\kappa_{\mathrm{grey}} = 0.1 \ \mathrm{cm}^{2} \, \mathrm{g}^{-1}$, the minimum opacity found in the analysis by \citet{Tanaka2019}.
\begin{figure}
    \centering
    \includegraphics[width=1\columnwidth]{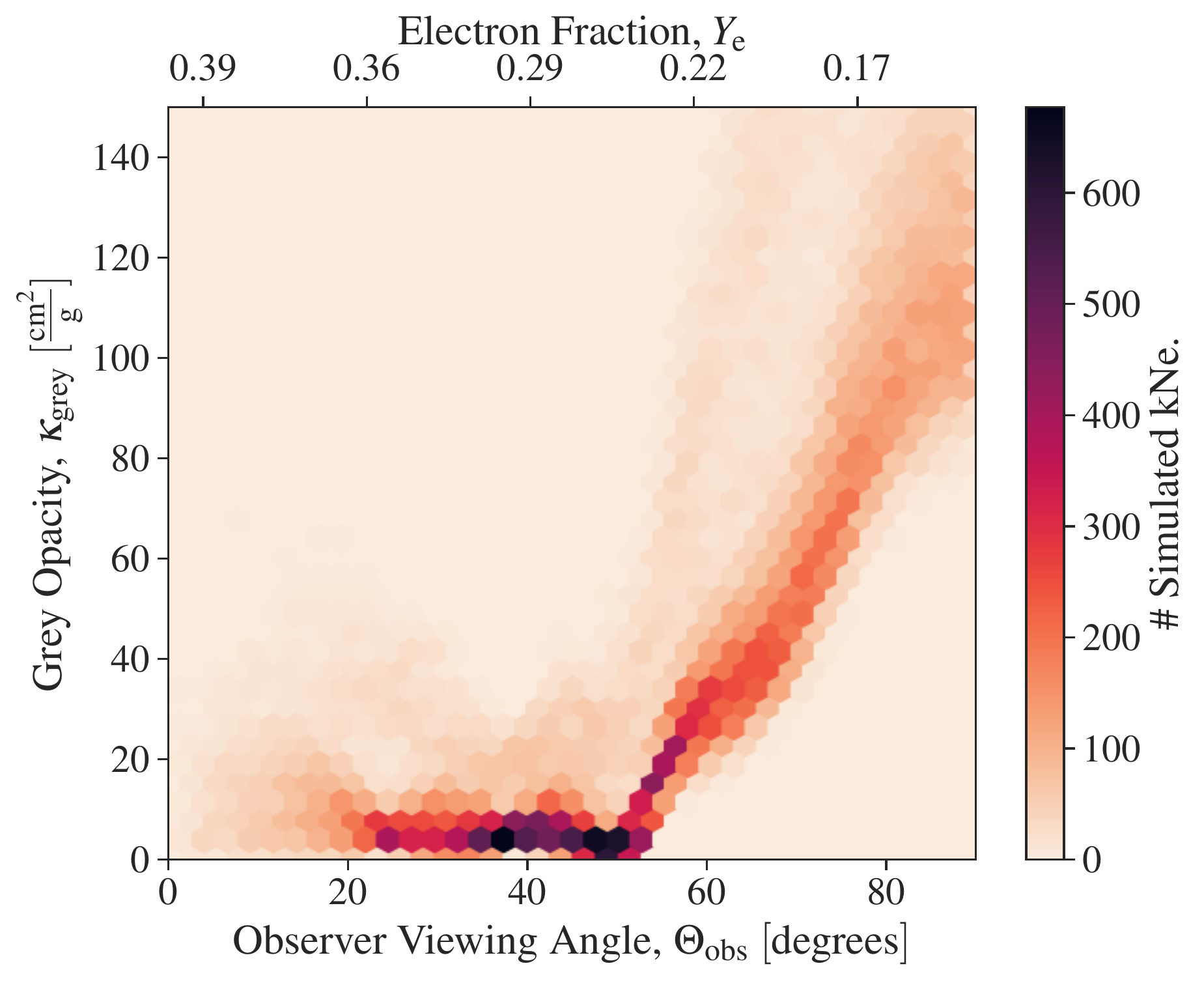}
    \caption{2D histogram of the population of BNS kNe viewed in the predicted $\kappa_\mathrm{grey} - \Theta_\mathrm{obs}$ plane. The viewing-angle, and hence the composition, has a strong relationship with the grey opacity seen most clearly at viewing-angles greater than $55$ deg., i.e., when viewed closer to edge-on. This transition to higher opacities implies an upper-limit opening angle of the lanthanide-rich ejecta of this population of approximately 35 deg. about the merger plane.}
    \label{Fig:GP_kgrey_viewing}
\end{figure}
We generally find the transitional behaviour of the grey opacity at an electron fraction of $Y_\mathrm{e} = 0.2-0.25$, see Fig. \ref{Fig:GP_kgrey_viewing}. Studies of this mapping generally agree that there is a transition in the opacity, due to the change in elements abundances of the material, around an electron fraction of $Y_{\mathrm{e}} \simeq 0.25$ \citep{Korobkin2012, Kasen2013, Lippuner2015}, also see fig. 2 of \citet{Rosswog2018}. \par 

\begin{figure}
    \hspace{-0.5cm}
    \centering
    \includegraphics[width=\columnwidth]{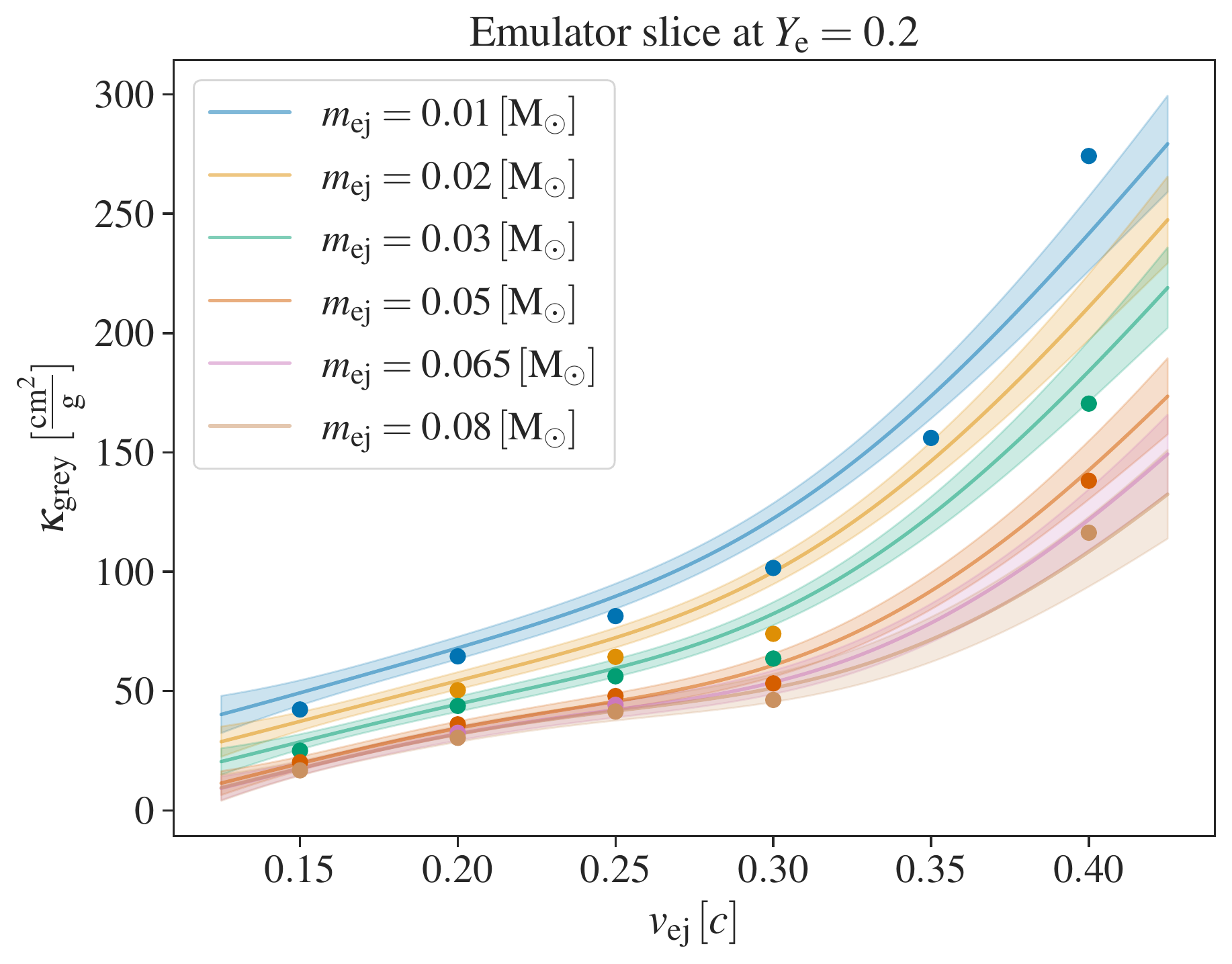}
    \caption{Slice through the ejecta parameter space of the Gaussian process emulator for the kN grey opacity. We show a slice through the region of ejecta parameter space predicted by the mappings of our population prior. This slice illustrates the dependence of the grey opacity on the median ejecta velocity of the material. This view of the emulator is shown at a fixed electron fraction, i.e., $Y_\mathrm{e} =0.2$ near the transition region, stacking the solutions for each of the training data values of ejecta mass.}
    \label{Fig:GP_param_slice}
\end{figure}

We find that the range of opacity values predicted for any electron fraction is mildly dependent on the ejecta mass and velocity of the material, see Fig. \ref{Fig:GP_param_slice}. As the grey opacity is the only free parameter in the emulator we see that some of this dimming behavior is reflected in the emulated grey opacity, such that lower total ejecta mass and higher median ejecta velocity lead to a larger grey opacity. This arises as our SAEE kN model does not simulate the same detailed physics as {\sc SuperNu} (such as the density-dependent thermalisation prescription). Thus, as we are fixing all parameters apart from the grey opacity, the physical dependencies get pushed into the variation of the grey opacity values we infer. 

This completes the set of parameters necessary to simulate kN-signals with our SAEE model. Given the parameter mappings above, we can directly generate a grey-body spectral-timeseries given the source-frame neutron star component masses, observer viewing-angle, and choice of EOS. \footnote{This model will be available as a standalone {\it pip}-install-able package, with documentation, available at: \url{https://github.com/cnsetzer/Setzer2022_BNSpopkNe}, and will be updated as improvements are made.}

\section{Results \& Discussion}\label{sec: results}
We have simulated a population of BNS kNe from realistic priors, incorporating EOS information, viewing-angle dependence, and relations between the intrinsic binary parameters and resulting kNe ejecta. This has allowed us to construct a population model of kNe light curves consistent with a progenitor BNS population.
\begin{figure}
    \centering
    \includegraphics[width=1\columnwidth]{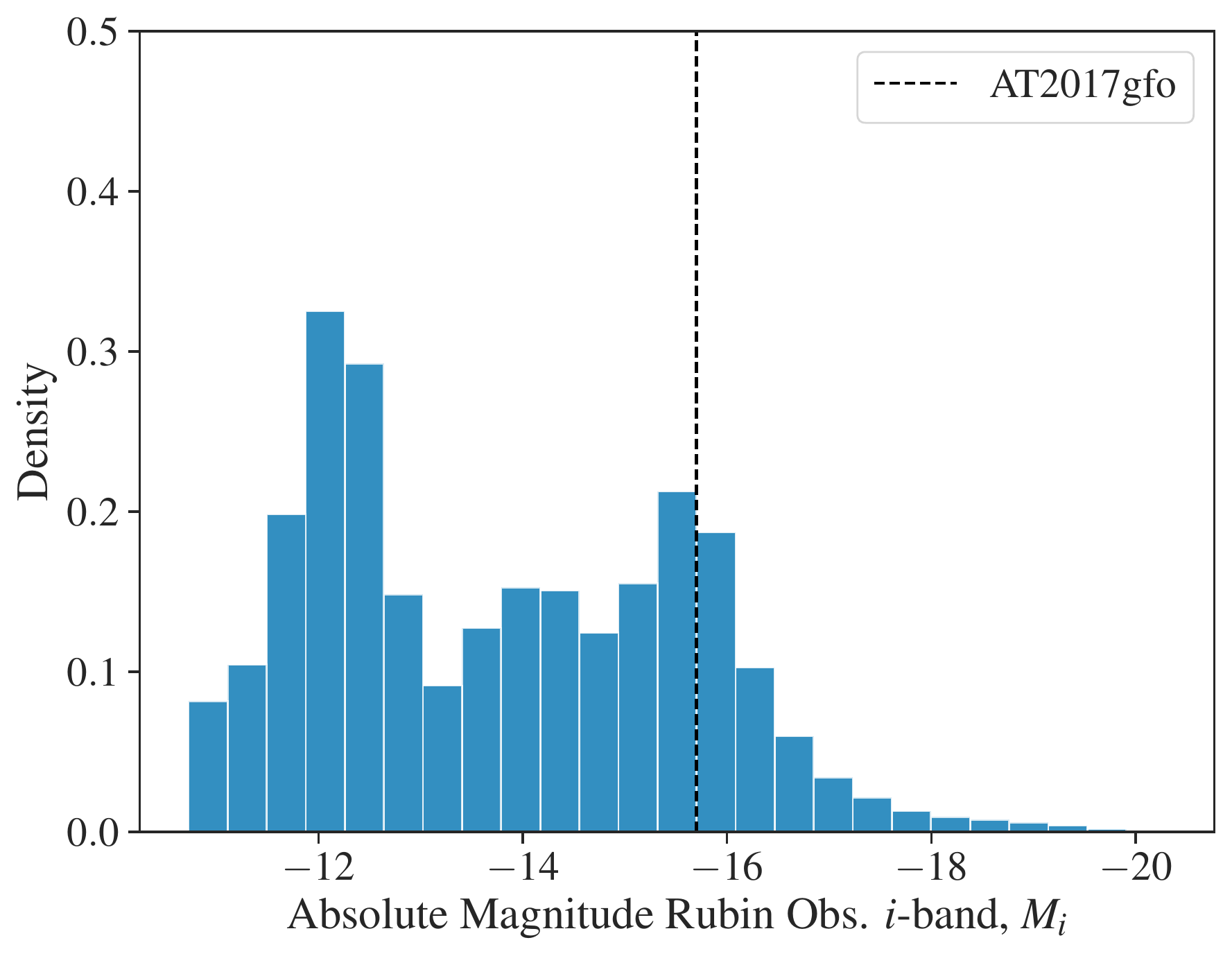}
    \caption{Peak absolute magnitude in the LSST {\it i}-band for a population of BNS merger kNe. The population has two peaks in the distribution which correspond to kNe which are either lanthanide-free along the observer line-of-sight, i.e., low electron fraction and thus brighter transient, or lanthanide-rich in the direction of the observer, i.e., high electron fraction leading to more opaque material and a dimmer transient. The vertical line indicates the peak {\it i}-band absolute magnitude of GW170817/AT2017gfo as inferred from observations, i.e., $M_i\, = \, -15.7$ from \citet{Smartt2017}.}
    \label{Fig:abs_mag_dist_i}
\end{figure}
\subsection{Characteristics of the Simulated kN Population}
\begin{figure}
    \centering
    \includegraphics[width=1\columnwidth]{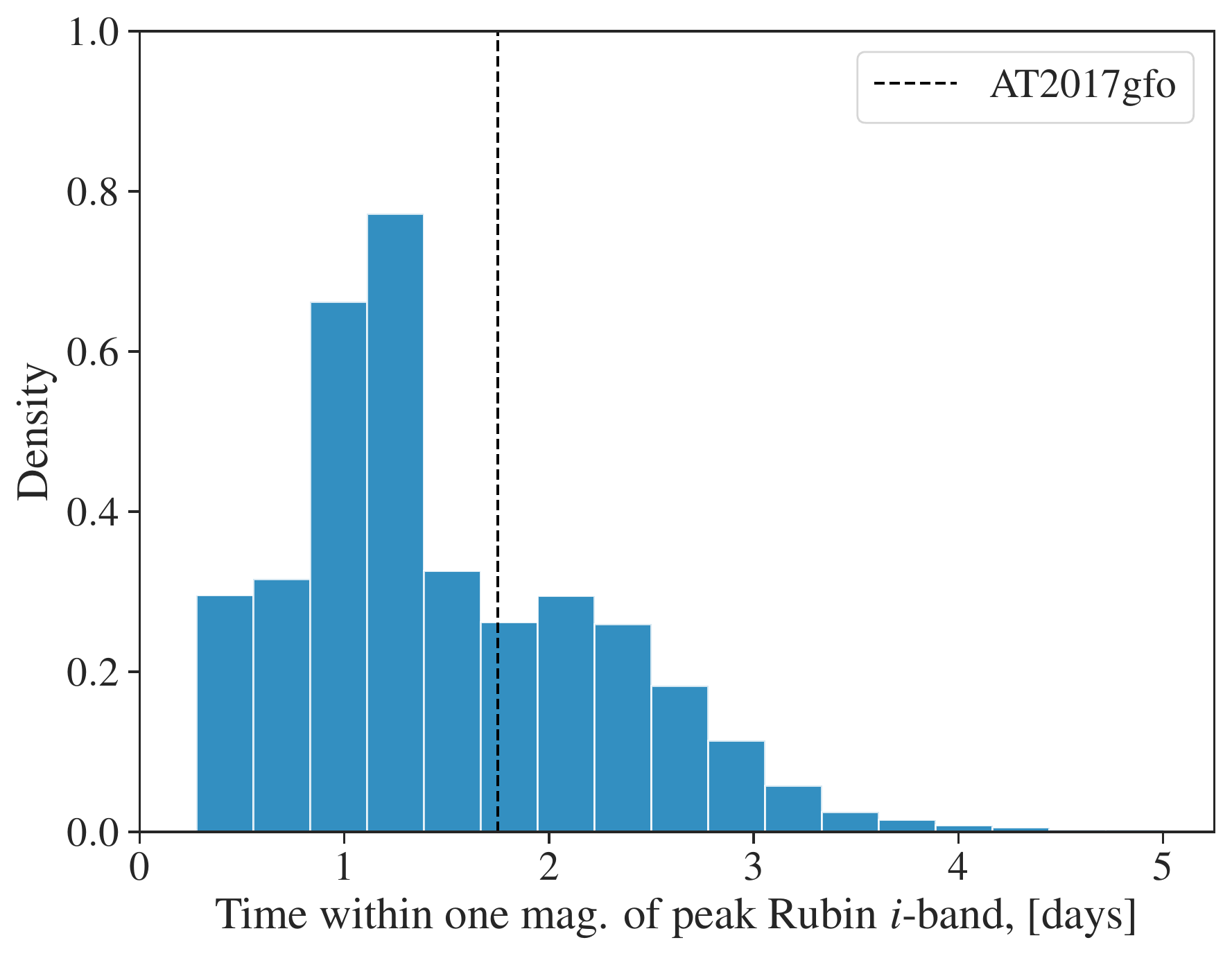}
    \caption{Distribution of time spent within one magnitude of the peak brightness in the LSST {\it i}-band by our simulated population. The distribution peaks around 1.25 days, indicating that the majority of BNS kNe spend around $0.75-3$ days within one magnitude of the peak brightness.}
    \label{Fig:brightness_timescale_i}
\end{figure}

Given the complex distribution of the sampled kN-parameters, as illustrated in Fig. \ref{Fig:dynamical_ejecta}, we expect a rich diversity of light curves predicted by this population. This is seen in the distribution of peak magnitudes, Fig. \ref{Fig:abs_mag_dist_i}, and also the distribution of the duration when the light curve is within one magnitude of the peak magnitude, see Fig. \ref{Fig:brightness_timescale_i}. As models of kNe generally exhibit their maximum peak magnitudes in the redder optical/near-infrared wavelength range, we show these distributions in the Rubin Observatory's {\it i}-band. 
We find the peak brightness for the population varies between $-20 \leq M_i \leq -11$. The distribution is approximately bimodal, with central peak-magnitudes of approximately $-15.5$ and $-12$ respectively. The brighter peak is correlated with lanthanide-free, i.e., high electron-fraction, low-opacity kNe and the dimmer peak with lanthanide-rich, i.e., low electron-fraction, high-opacity kNe. We also show in Fig. \ref{Fig:abs_mag_dist_i} that the peak magnitude of GW170817/AT2017gfo is compatible with our kNe population. The distribution of time spent near peak brightness in this band peaks around $1.25$ days with a minimum duration of approximately $0.25$ days and a long tail to higher values. We note this tail is comprised of high-opacity kNe with slow-moving ejecta. Although the long time-scales would be optimistic for detection, these kNe are inherently the dimmest population, and modelling of this part of parameter space is more uncertain. 
\begin{figure*}
  \centering
  \includegraphics[width=\textwidth]{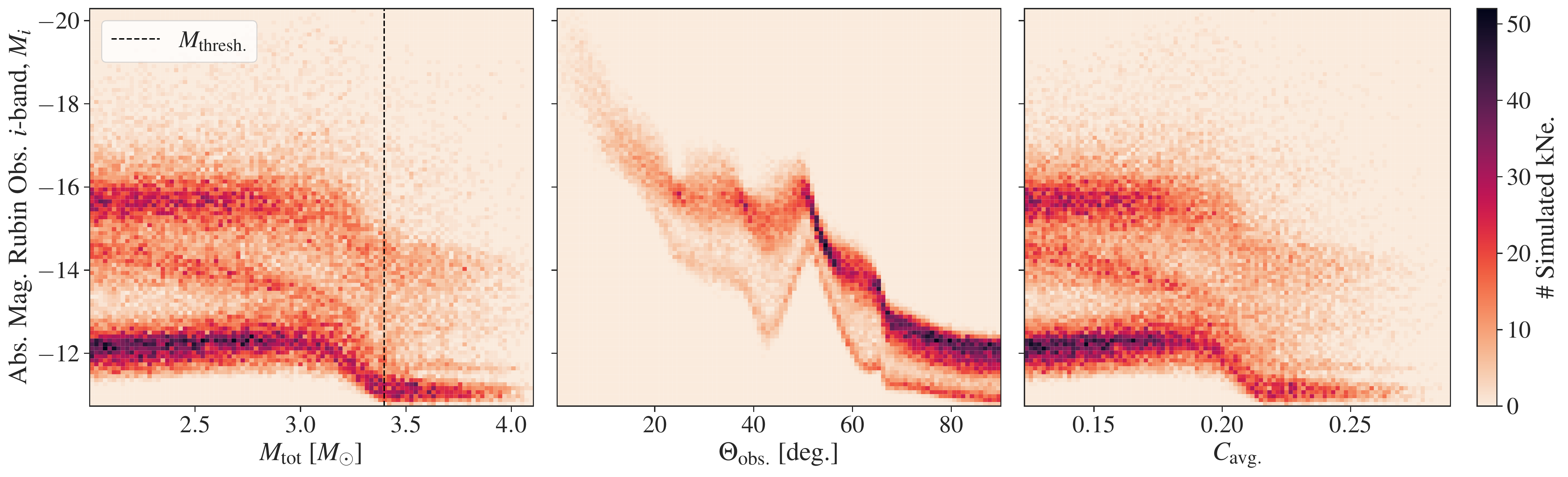}
  \caption{Two-dimensional histograms showing the variation of Rubin Obs. {\it i}-band absolute magnitude against intrinsic parameters of the binaries we simulate. (Left) Variation w.r.t the total mass of the binary, the vertical line indicates the threshold mass for prompt black hole collapse. (Middle) The variation w.r.t. the observer viewing-angle. (Right) The variation w.r.t. the compactness of the neutron stars averaged over the two components of the binary.}
  \label{Fig:intrinsic_variation}
\end{figure*}

\begin{figure}
    \centering
    \includegraphics[width=\columnwidth]{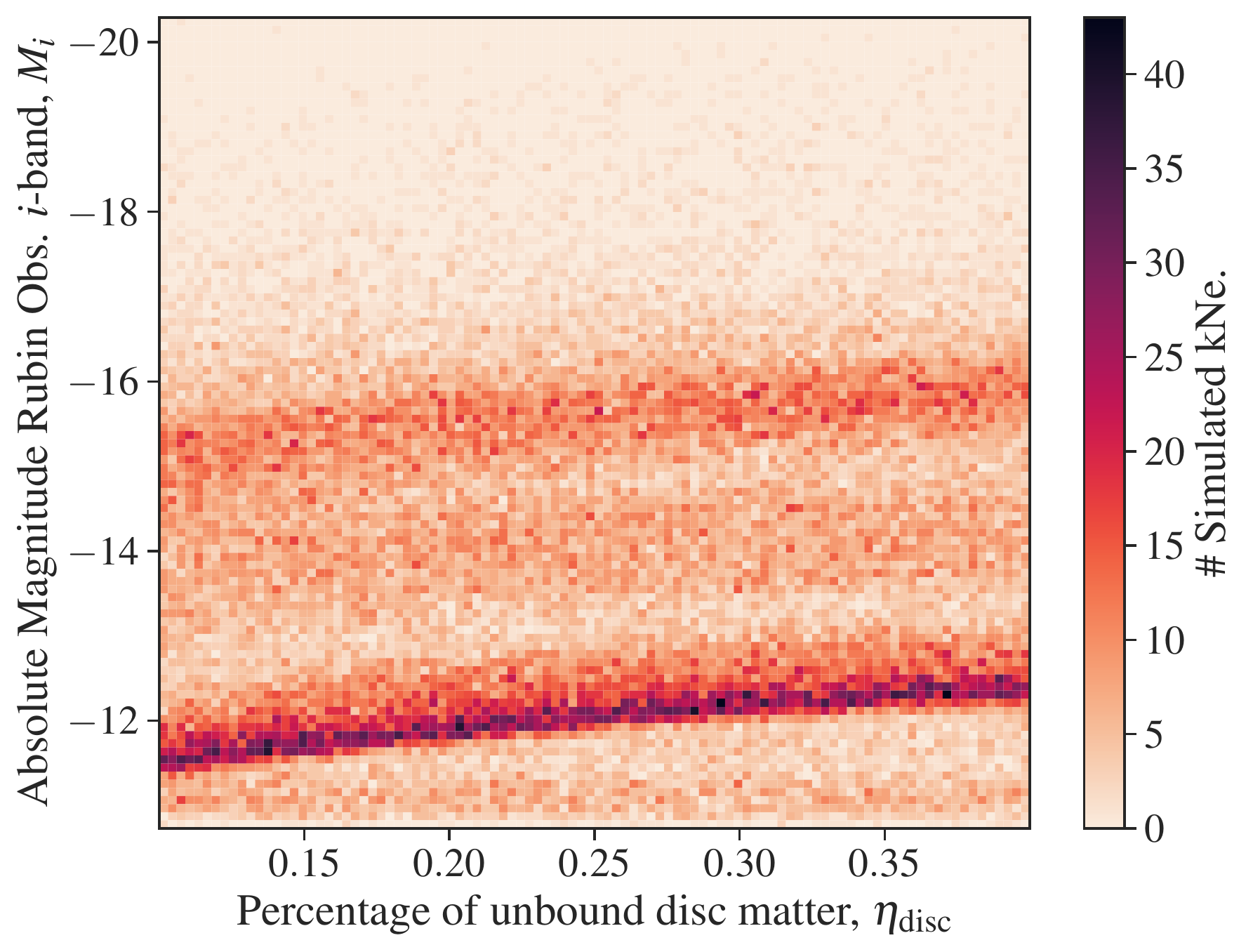}
    \caption{Two-dimensional histogram of the variation in {\it i}-band peak brightness vs. the fraction of material unbound from the remnant disc. Considering the mean of the two populations coming from high/low opacities, we see the percentage of unbound disc matter contributes at most an approximate 0.5 magnitude change in the resulting peak brightness.}
    \label{fig:disk_unbinding}
\end{figure}
We see in Fig. \ref{Fig:intrinsic_variation} that the population peak brightness strongly depends on the viewing-angle of the binary. There is at least a three magnitude difference from face-on to edge-on orientations. This is expected due to the deterministic relationship of the viewing-angle to the electron fraction composition and the strong relationship between electron fraction and grey opacity, see Fig. \ref{Fig:GP_kgrey_viewing}. This illustrates that the peak brightness is very sensitive to the composition. We have considered the case of the mass-weighted average composition profile, with contributions only of the line-of-sight material. It will be important to investigate in future work the impact of non-line-of-sight components of varying composition, aspherical morphologies, and the effect of differences in composition between the inner and outer regions of the ejecta. Studies which have already considered some of these effects, such as \citet{bulla_possis_2019}, predict peak brightnesses within the range predicted by our population.

In the same figure, we also see a mild trend to greater peak brightness with increasing total mass. However, once the total mass approaches the threshold for prompt black hole collapse, there is a sharp decrease of about 1 magnitude in the peak brightness. In the middle panel we see some modelling artefacts related to the Gaussian process interpolation in the range around $50$ degrees. It is very difficult to model the sharp change in opacity in this region without a large increase in training data. The artefact is also related to the piece-wise prescription of the mean function specified in our grey-opacity fitting scheme, see Sec. \ref{sec:GP_emul}. 

A significant modelling uncertainty is the contribution to the total ejecta mass coming from the amount of matter unbound from the remnant accretion disc. In Fig. \ref{fig:disk_unbinding}, we show the variation in peak brightness due to the modelling uncertainty of the disc unbinding-fraction. Over the range of unbinding percentages considered, we see an approximate 0.5 mag change in the peak brightness. This clearly shows that the unbinding uncertainty is subdominant to the viewing-angle contribution to peak brightness. Indeed, we find the dependence of peak brightness with respect to all other parameters of the model is subdominant to the viewing-angle/composition contribution.

\subsection{Comparison to GW170817/AT2017gfo}
We find that predictions from our population model are able to reproduce features observed in the light curves of AT2017gfo, despite not calibrating the model to this event. Specifically, we see that the population is able to produce kNe with the same peak brightness and peak duration as GW170817/AT2017gfo.  

\begin{figure*}
    \centering
    \includegraphics[width=1\textwidth]{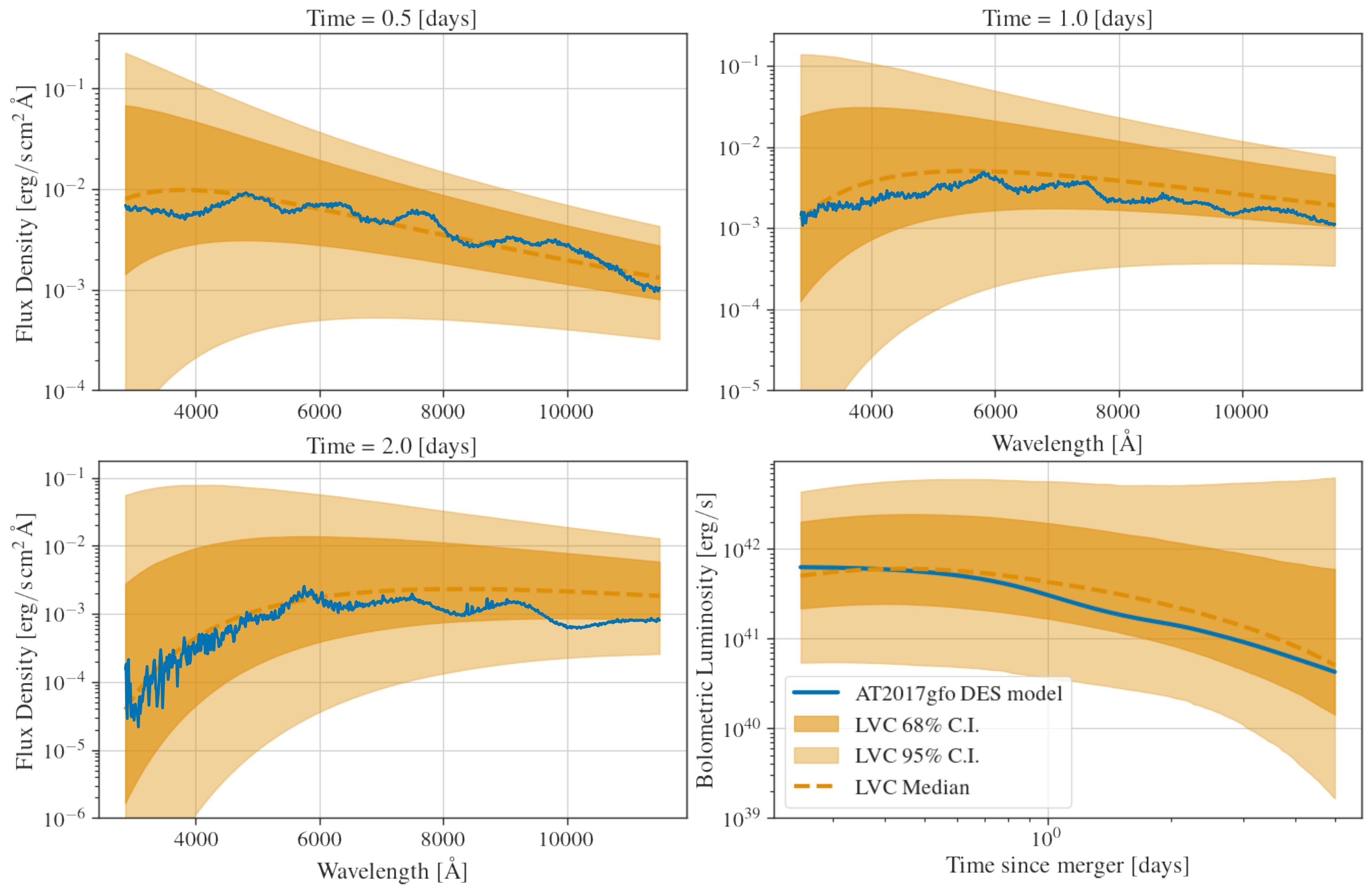}
    \caption{Similar to Fig. \ref{Fig:model_mistmatch}. This compares the DES spectral time series model of AT2017gfo \citep{Scolnic2017a} to the range of kNe produced with the SAEE model for 5000 draws of the component masses and viewing angle from the LVC posterior for GW170817 \citep{GW170817props}. These binary parameters are mapped to the associated kN ejecta parameters through the relations specified in Sec. \ref{sec:kn_model}. Though our kN model does not include spin effects, we draw samples from the LVC high-spin posterior due to its greater support for higher mass ratios compatible with the full range we consider when simulating the population of kNe produced in this work.}
    \label{Fig:GW170817_comparison}
\end{figure*}

Considering Fig. \ref{Fig:GW170817_comparison}, we see that the smooth spectrum of our grey-body model captures the overall spectral shape of the observations-based model of AT2017gfo \citep{Scolnic2017a}. To make this comparison we did not fit the SAEE model to observations of AT2017gfo, but rather sampled the component masses and viewing-angle from the binary parameter posteriors for GW170817 \citep{GW170817props} to construct a posterior predictive test for AT2017gfo using our model. Given the GW170817 posterior, our model predicts a range of kNe from which we construct the quantiles shown in Fig. \ref{Fig:GW170817_comparison}. For binary parameters consistent with the gravitational wave signal, our model produces kNe that show a $2$--$3$ orders of magnitude variation in the amplitude of the flux and luminosity, which is due to the wide range of mass-ratios and inclinations supported by the GW170817 posterior. We find that the median spectra are in good agreement with observations-calibrated modelling of the kNe emission from AT2017gfo \citep{Scolnic2017a}.

The median total luminosity over the observationally relevant region is also in close agreement with predictions of the DES GW170817 spectral model \citep{Scolnic2017a}. The grey-body model captures the time-scales of the smooth rise and fall of the luminosity. These results demonstrate overall self-consistency of our model with the gravitational wave and electromagnetic data on this single event; however, please note that one should not expect to find detailed agreement with second-order effects in time-evolution with the single-component grey-body modelling used in this work.

Our results indicate overall that detection prospects for kNe populations derived assuming that GW170817/AT2017gfo is typical are likely too optimistic. While our population can accommodate the features of GW170817, it is brighter than the typical kN in our population. However, with only one BNS kN confirmed with multi-messenger observations, we stress that the uncertainties in modelling make it difficult to draw definitive conclusions on the prospects of future detections. It is clear that observing the next kNe will be crucial to inform modelling the broader population of kNe.
\section{Conclusions}\label{sec: conclusion}
We have modelled a population of kNe by relating the parameters of neutron star binaries to the parameters of kN ejecta, enabling simulation of light curves dependent only on the properties of the binary. This utilizes a series of mappings constructed from numerical simulations of more detailed physics. Specifically, we have constructed an analytic description of the composition profile for the outermost ejecta. Using detailed radiation transport simulations we have calibrated a grey opacity model and trained a Gaussian process emulator to predict grey-opacities over a broad parameter space of kN ejecta.

We find that the resulting population of BNS merger kNe produce a diverse range of kN-signals. Considering the most observationally relevant regions of the kN evolution, we predict a range of peak brightness, measured as the peak absolute magnitude in the Rubin observatory {\it i}-band, to be in the range of $-20 \leq M_i \leq -11$. We additionally find that kNe are indeed short-lived transients, with the duration of their light curves around peak brightness lasting $\sim1.25$ days. Our results are consistent with the cosmological kNe population model presented by \citet{Colombo2022}. Both analyses predict a short-lived population of kNe, spanning at least six magnitudes in peak brightness, and peak-times lasting a few days at most, see fig. 2 and fig. 10 of \citet{Colombo2022}.

Our population model is consistent with GW170817, reproducing key features of the observations. The model also predicts a bimodal light curve distribution due to the transition from high to low electron fraction ejecta, which supports the commonly used assumption of a binary opacity choice when simulating kN light curves. However, it is clear that grey-body modelling has limitations, particularly in reproducing the detailed spectral shape. Furthermore, as we have our calibrated the model to early-time, near-peak magnitudes of the light curves, there is significant uncertainty in the late-time predictions of the model.

This work represents a major step toward developing a robust population model for kNe. Better linking of BNS properties to kNe properties, via empirical mappings calibrated with simulation data, will be key to building further on this work. The connection of the light curve to the intrinsic parameters of the BNS enables studies of the combined EM-GW observational prospects of such a population. Additionally, observational prospects for different binary population synthesis models or equations of state can be explored. Furthermore, light curve simulations dependent on the parameters of the binary makes possible a joint analysis of GW and kN data, where both signals are sampled from a self-consistent population prior. In future work, we will use this population model to derive the observational selection function for EM-GW events, which is a critical ingredient for cosmological inference with such populations.

\section*{Acknowledgements}
We thank David Radice for discussions regarding use of the simulation data from which we derived the fits for the electron fraction. We also thank Keir Rogers regarding for discussing efficient ways for optimising narrow likelihoods in large parameter spaces. Further we thank Masaomi Tanaka for discussing an appropriate lowest bound on kN grey opacities. We are grateful to Nikhil Sarin for helpful feedback on the manuscript and for conducting software review on the emulator code release. This research utilised the {\it Sunrise} HPC facility, supported by the Technical Division at the Department of Physics, Stockholm University. This research made use of {\sc Astropy},\footnote{\url{http://www.astropy.org}} a community-developed core Python package for Astronomy \citep{astropy:2013, astropy:2018}. This research also made use of {\sc SNCosmo} to manage SEDs of simulated kNe \citep{Barbary2014}. This work additionally has used the Python software packages {\sc Numpy} \citep{harris2020array}, {\sc Scipy} \citep{2020SciPy-NMeth}, and {\sc Pandas} \citep{mckinney-proc-scipy-2010, Reback2020}.

CNS acknowledges travel funding provided by the LSST Corporation. HVP was partially supported by the research project grant “Fundamental physics from populations of compact object mergers” funded by VR under Dnr 2021-04195. The work of HVP was additionally supported by the Göran Gustafsson Foundation for Research in Natural Sciences and Medicine. This project has received funding from the European Research Council (ERC) under the European Union’s Horizon 2020 research and innovation programme (grant agreement no. 101018897 CosmicExplorer). This work has been enabled by support from the research project grant ‘Understanding the Dynamic Universe’ funded by the Knut and Alice Wallenberg Foundation under Dnr KAW 2018.0067. SR has been supported by the Swedish Research Council (VR) under grant number 2020-05044, by the research environment grant “Gravitational Radiation and Electromagnetic Astrophysical Transients” (GREAT) funded by VR under Dnr 2016-06012, by the Knut and Alice Wallenberg Foundation under grant Dnr. KAW 2019.0112, by the Deutsche Forschungsgemeinschaft (DFG, German Research Foundation) under Germany’s Excellence Strategy – EXC 2121 “Quantum Universe” – 390833306 and by the European Research Council (ERC) Advanced Grant INSPIRATION under the European Union’s Horizon 2020 research and innovation programme (Grant agreement No. 101053985) This work is supported by the LANL ASC Program and LDRD grants 20200145ER and 20190021DR. This work used resources provided by the LANL Institutional Computing Program. LANL is operated by Triad National Security, LLC, for the National Nuclear Security Administration of the U.S.DOE  (Contract No. 89233218CNA000001).

The contributions from the authors are listed below: {\bf C.N.S.}: conceptualisation; data curation; formal analysis; investigation; methodology; project administration; software; validation; visualization; writing - original draft; writing - review \& editing. {\bf H.V.P.}: conceptualisation; formal analysis; funding acquisition; methodology; investigation; project administration; supervision; validation; writing - review \& editing. {\bf O.K} data curation; resources; software; investigation; validation; writing - review \& editing {\bf S.R.}: resources; software;  writing - review \& editing.

\section*{Data availability}
The data underlying this article will be shared on reasonable request to the corresponding author. The model used to generate all data for the results will be available at \url{https://github.com/cnsetzer/Setzer2022_BNSpopkNe}.



\bibliographystyle{mnras}
\bibliography{kne_paper2_refs}




\appendix
\section{Derivation of Density-averaged Time-dependent Thermalization Efficiency}\label{sec:appendix}
Here we present the derivation of a density-averaged form of the thermalisation efficiencies used in {\sc SuperNu} as specified by \citet{Wollaeger2017}. The total heating rate, $\dot{Q}_i(t)$ [erg/s], due to species ``$i$'', either $\alpha$, $\beta$, or $\gamma$ radiation, and fission fragments, can be specified by integrating the specific heating rate, denoted by $\dot{\varepsilon}$ [erg/s/g], over the full ejecta outflow,
\begin{equation}\label{eq:int_tot_eng}
    \dot{Q}_i(t) = 4\pi \int_0^R r^2dr \cdot \rho(r,t) f_i(r,t) \dot{\varepsilon}_i(t), 
\end{equation}
where $r$ is the radial coordinate of the ejecta, $R$ is the leading edge of the outflow, $\rho(r,t)$ is the density of the ejecta, and $f_i(t)$ is the efficiency of thermalisation for a given species.

For $\alpha$ and $\beta$ radiation and fission fragments, the heating efficiency is specified, following \citet{barnes_radioactivity_2016} by,
\begin{equation}\label{eq:therm_eff}
    f_i(r,t) = \frac{\log \left(1 + 2\eta_i^2(r,t) \right)}{2\eta_i^2(r,t)},
\end{equation}
where
\begin{equation}\label{eq:2eta}
    2\eta_i^2(r,t) = \frac{2A_i}{t\rho(r,t)},
\end{equation}
and $A_i$ are thermalisation time constants: $\{A_\alpha, A_\beta, A_\mathrm{ff}\} = \{1.2, 1.3, 0.2\} \times 10^{-11} \mathrm{g\ cm}^{-3}\mathrm{s} $. Substituting these expressions into Eq. \ref{eq:int_tot_eng}, and utilizing the homologous expansion approximation where the expansion velocity is $v = r/t$, and performing variable substitution $x=vt$,
\begin{align}
    \dot{Q}_i(t) &= 4\pi \dot{\varepsilon}_i \int_0^{v_\mathrm{max}} (vt)^2 d(vt) \cdot \rho \cdot \frac{t \rho}{2A_i} \log \left(1+ \frac{2A_i}{t\rho} \right), \\
    &= 4\pi \dot{\varepsilon}_i \frac{t^2 v_\mathrm{max}^3}{2A_i} \int_0^1 x^2dx(t\rho)^2 \log \left(1+ \frac{2A_i}{t\rho} \right). \label{eq:int_q_3}
\end{align}
We can then utilize the spherically-symmetric density profile used by {\sc SuperNu} \citep{Wollaeger2017}:
\begin{equation}\label{eq:spherical_density}
    \rho(r,t) = \rho_0 \left(\frac{t}{t_0} \right)^{-3} \left(1 - \frac{v^2}{v_\mathrm{max}^2} \right)^3,
\end{equation}
where $t_0$ is a reference time and $\rho_0$ is the central density at the reference time. Given the ejecta mass, $m_\mathrm{ej}$, and the maximum expansion velocity, $v_\mathrm{max}$, the reference central density can be expressed in terms of the ejecta parameters,
\begin{equation}\label{eq:ref_rho}
    \rho_0 = \frac{315}{64\pi}m_\mathrm{ej}(v_\mathrm{max}t_0)^{-3}.
\end{equation}
By introducing dimensionless time-dependent constants
\begin{equation}\label{eq:alpha}
\alpha_i = \frac{2A_i}{t \rho_0}\left( \frac{t}{t_0}\right)^3,
\end{equation}
and substituting the density profile, Eq. \ref{eq:spherical_density}, into Eq. \ref{eq:int_q_3}, the total ejecta heating rate becomes:
\begin{align}\label{eqn:integral_edot}
    \dot{Q}_i(t) &= 8\pi \dot{\varepsilon}_i(t) \frac{A_i t^2 v_\mathrm{max}^3}{\alpha_i} \times \\ &\left[ \frac{1}{\alpha_i} \int_0^1 x^2dx \cdot (1-x^2)^6 \log\{1+\alpha_i(1-x^2)^{-3} \}  \right ]. \nonumber
\end{align}
The bracketed expression in Eq. \ref{eqn:integral_edot} can be approximated as
\begin{equation}
    I(\alpha) \approx B \frac{\log(1+\mathrm{c}\alpha)}{\mathrm{c}\alpha},
\end{equation}
where $B = 0.0507935$ and $\mathrm{c}=2.3$ are the fit constants found by equating this expression to the numerical integration of this quantity from {\sc SuperNu}. 

Substituting this back into Eq. \ref{eqn:integral_edot} we arrive at the following expression for the total heating rate of the ejecta due to species ``$i$'',
\begin{equation}\label{eq:approx_qdot}
    \dot{Q}_i(t) \approx  8\pi B \dot{\varepsilon}_i(t) \frac{A_i t^2 v_\mathrm{max}^3}{\alpha_i} \frac{\log(1+\mathrm{c}\alpha_i)}{\mathrm{c}\alpha_i}.
\end{equation}
Now, considering the semi-analytic model, which assumes the heating is independent of the density of the ejecta, the total heating rate of the ejecta can be approximated as
\begin{equation}\label{eq:saee_qdot}
    \dot{Q}_i(t) \approx m_\mathrm{ej} \dot{\varepsilon}_i(t) \bar{f}_i(t),
\end{equation}
where $\bar{f}_i(t)$ is an averaged thermalisation efficiency. If we rewrite Eq. \ref{eq:alpha} for $\alpha_i$ in terms of the ejecta parameters, substituting in Eq. \ref{eq:ref_rho},
\begin{equation}\label{eq:alpha2}
    \alpha_i = \frac{128\pi}{315} A_i \frac{v_\mathrm{max}^3 t^2}{m_\mathrm{ej}},
\end{equation}
we can set Eq. \ref{eq:approx_qdot} equal to Eq. \ref{eq:saee_qdot} and solve for the averaged thermalisation efficiency. This leads to:
\begin{equation}\label{eq:fbar}
    \bar{f}_i(t) = \frac{315B}{16} \frac{\log (1 + c\alpha_i)}{c\alpha_i} \approx \frac{\log (1 + c\alpha_i)}{c\alpha_i},
\end{equation}
since $315B/16 \approx 1$. Noting that this is identical in functional form to Eq. \ref{eq:therm_eff}, we make the following equivalence:
\begin{equation}\label{eq:alpha_equiv}
c\alpha_i \equiv 2\bar{\eta}_i^2,
\end{equation}
where $2\bar{\eta}_i^2$ is a density-averaged form of Eq. \ref{eq:2eta}, i.e.,
\begin{equation}\label{eq:bar_eta}
    2\bar{\eta}_i^2 = \frac{2A_i}{t \bar{\rho}(t)}.
\end{equation}
Lastly, we can express the average density in terms of the ejecta parameters by equating Eq. \ref{eq:alpha_equiv} and Eq. \ref{eq:bar_eta} and substituting in Eq. \ref{eq:alpha2},
\begin{equation}\label{eq:rho_bar}
    \bar{\rho}(t) = \frac{16\pi B}{c}\frac{m_\mathrm{ej}}{(v_\mathrm{max}t)^3} = \frac{2\pi B}{\mathrm{c}}\frac{m_\mathrm{ej}}{(v_\mathrm{ej}t)^3}.
\end{equation}
In this we have approximated $v_\mathrm{ej} \approx (1/2) v_\mathrm{max}$ following \citet{Wollaeger2017}. Putting together Eq. \ref{eq:fbar} - Eq. \ref{eq:rho_bar} we find the density-averaged thermalisation efficiency for $\alpha$, $\beta$, and fission fragments is given by:
\begin{equation}
    \bar{f}_i(t) = \frac{\log \left(1+\frac{c A_i t^2 v_\mathrm{ej}^3}{\pi B m_\mathrm{ej}} \right)}{\frac{c A_i t^2 v_\mathrm{ej}^3}{\pi B m_\mathrm{ej}}}.
\end{equation}
For $\gamma$-rays we assume a thermalisation efficiency following \citet{Kasen2018a} and \citet{Wollaeger2017} given by
\begin{equation}
    \bar{f}_\gamma(t) = 1-e^{-\bar{\tau}(t)},
\end{equation}
where the $\gamma$-ray thermalisation time-scale is given by
\begin{equation}
    \bar{\tau}(t) = 0.035 \frac{\kappa_\gamma m_\mathrm{ej}}{(v_\mathrm{ej}t)^2},
\end{equation}
with $\kappa_\gamma =0.1 \ \mathrm{cm}^2\mathrm{g}^{-1}$.

\bsp	
\label{lastpage}
\end{document}